\documentclass[10pt]{article}
\usepackage{amsmath, amssymb, amsthm}
\usepackage{graphicx}
\usepackage{pstricks, pst-plot}
\usepackage{times, mathrsfs}
\usepackage{enumerate}
\usepackage{multirow,hhline}
\usepackage{setspace}
\usepackage{caption,subcaption}
\usepackage{authblk}

\usepackage[square]{natbib}
\begin{document}
\title{Mean-dependent nonstationary spatial models}
\author[1]{G.C.L. Peterson}
\author[1]{J. Guinness}
\author[2,3]{A. Terando}
\author[1,4]{B.J. Reich}
\affil[1]{Department of Statistics, North Carolina State University, Raleigh, United States}
\affil[2]{U.S. Geological Survey, Department of the Interior Southeast Climate Science Center, Raleigh, United States}
\affil[3]{Department of Applied Ecology, North Carolina State University, Raleigh, United States}
\affil[4]{Correspondance: Dr.G.C.L. Peterson, North Carolina State University, Department of Statistics, SAS Hall, Raleigh, 27695, USA, Email: geoffrey.colin.peterson@gmail.com}

\doublespacing
\maketitle
\begin{abstract}
Nonstationarity is a major challenge in analyzing spatial data. For example, daily precipitation measurements may have increased variability and decreased spatial smoothness in areas with high mean rainfall. Common nonstationary covariance models introduce parameters specific to each location, giving a highly-parameterized model which is difficult to fit. We develop a nonstationary spatial model that uses the mean to determine the covariance in a region, resulting in a far simpler, albeit more specialized, model. We explore inferential and predictive properties of the model under various simulated data situations. We show that this model in certain circumstances improves predictions compared to a standard stationary spatial model. We further propose a computationally efficient approximation that has comparable predictive accuracy. We also develop a test for nonstationary data and show it reliably identifies nonstationarity. We apply these methods to daily precipitation in Puerto Rico.

\end{abstract}
\section{Introduction}
\begin{figure}[p]
\centering
\includegraphics[height=0.5\textwidth]{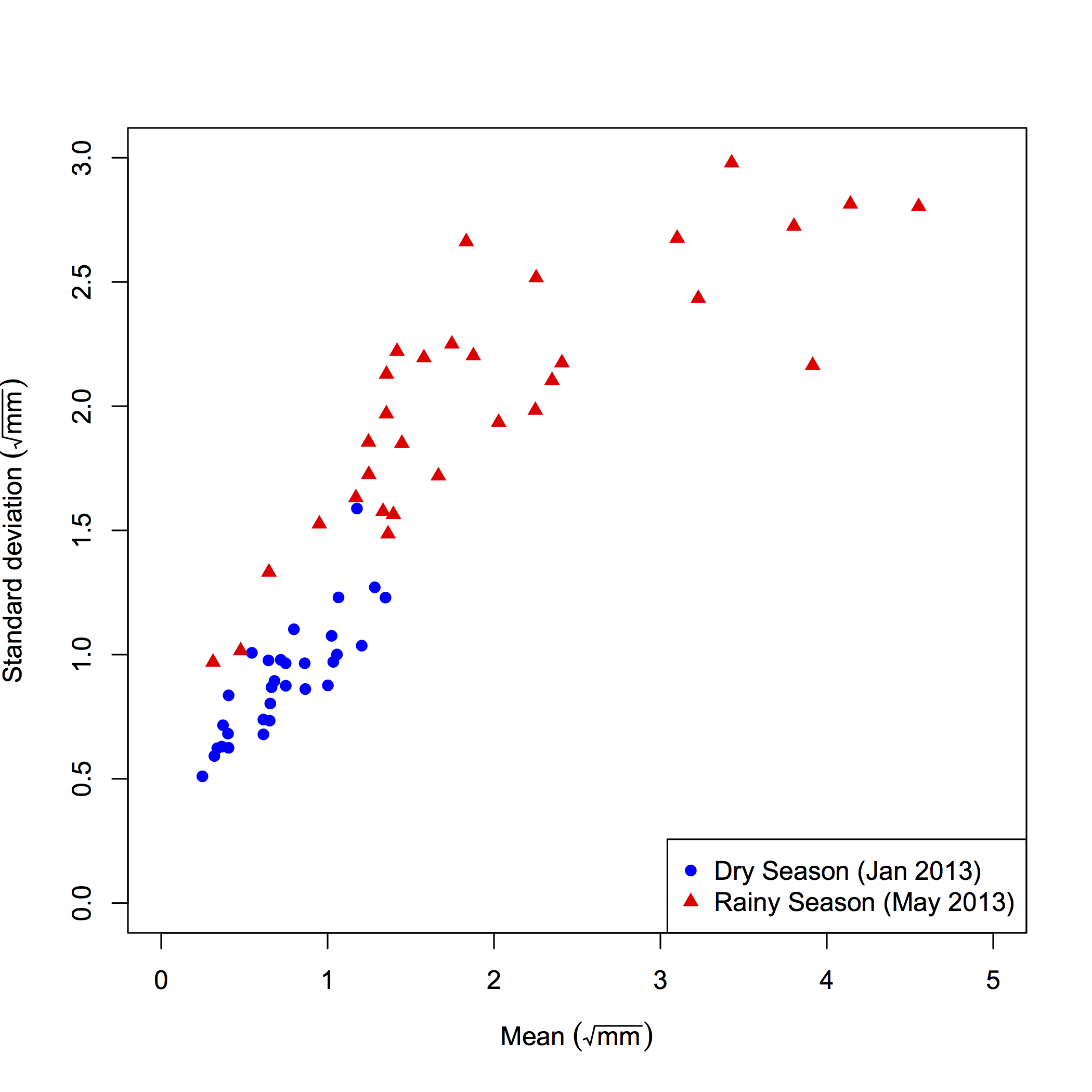}
\caption{Monthly means and standard deviations for Puerto Rico data. Rain gauge measurements (in millimeters) obtained from weather stations in Puerto Rico over January 2013 and May 2013. Data has been square-rooted.}
\label{fig:PR_station_mean_var}
\end{figure}

Modeling daily precipitation poses unique challenges to typical spatio-temporal models. Sites experiencing higher amounts of precipitation often exhibit larger variability than sites experiencing low amounts of precipitation, as seen in Figure \ref{fig:PR_station_mean_var}. Furthermore, sites in regions experiencing similar weather patterns will be more correlated than sites experiencing different weather patterns, so the spatial covariances will change over the spatial domain. Given that the most widely used long-term daily precipitation time series do not account for this, and as the regions that experience high precipitation changes daily, we need a flexible model that accounts for changes in the daily mean and spatial covariance structures. The mean structure for a given day can be modeled using specified covariates, but we need a model that allows the spatial covariances between sites to shift both from day to day and from region to region.  

Unlike stationary models, where the covariance structure is consistent across the spatial domain, nonstationary models allow for different levels of variability and spatial smoothness within the same domain. Nonstationary models have been developed through multiple approaches, which are summarized in \cite{sampson2010}. Typical techniques for nonstationary models are spatial deformation methods \citep{anderes2008,sampson1992, schmidt2003}, basis function models \citep{holland1999, nychka1998, nychka2002}, kernel based methods \citep{fuentes2001, fuentes2001tech, higdon1998, reich2013}, and spectral methods \citep{fuentes2002, guinness2015}.

Regime-switching models are another class of the nonstationary spatio-temporal covariance models. Regime-switching models use a latent process, called the spatial regime, to determine the mean and spatial covariance of each site in the domain \citep{paciorek2006}. Discrete regime models are common in hidden Markov models \citep{baum1966, barbu2006, leroux1992, lindgren1978}, econometrics \citep{gray1996, zhou2003}, wind forecasting \citep{gneiting2006, pinson2012, neto2014}, and wind time series \citep{shamshad2005, kazor2015}. In these cases, the underlying physics of the process switches between specific modes, such as bull versus bear markets or directional shifts of major wind currents. However, discrete regimes are not always flexible enough to account for geographic and temporal shifts in the size and shape of regimes.
 
Instead of discrete regimes, continuous regime processes allow for more complex patterns in the nonstationary spatial model. Several authors have already extended discrete spatial regimes into continuously-indexed spatial regimes. \cite{kleiber2012} extended the method for multivariate spatial processes, although without considering detailed modeling of the latent regime process. Modeling the regime process using covariates have been used extensively \citep{calder2008, reich2011, schmidt2011, risser2015}. These covariate-driven models create continuously-varying regimes that directly inform the spatial covariance structure. In particular, temporal covariates can flexibly account for each day's weather, whether it is clear skies, light rain, or even hurricane-level storms.

This paper builds on continuous regime-switching models but deviates from other approaches by indexing the regimes using the mean process. That is, we allow a separate spatial covariance for generally dry regions versus generally wet regions, but assume that all dry spatiotemporal regions have the same spatial covariance. While the proposed method cannot capture all forms of nonstationarity, it is useful in the specific case where mean/covariance relationship are thought to be critical. The data and the resulting spatial model is then used to create a predictive distribution across the entire spatial domain with appropriately varying standard errors. We discuss computation of both stationary and nonstationary models within this framework and develop a test statistic for significant departures from stationarity. We illustrate the model's predictive capabilities using daily precipitation data from Puerto Rico and confirm that as the mean precipitation in an area on a given day increases, the variability increases and spatial smoothness decreases.
\section{Nonstationary spatial model from conditionally stationary regimes} \label{sec:model}
Let $Y_t(\mathbf{s})$ be the response at spatial location $\mathbf{s}$ and time $t$. The data are observed at locations $\mathbf{s}_1,\dots,\mathbf{s}_n$ and days $t_1,\dots,t_m$. The response is decomposed as $Y_t(\mathbf{s}) = \mu_t(\mathbf{s}) + e_t(\mathbf{s})$, where $\mu_t(\cdot)$ is the mean process and $e_t(\cdot)$ is, conditionally on $\mu_t(\cdot)$, a zero-mean Gaussian error process. The mean function is modeled as a linear combination of $J$ known covariates $Z_1(\mathbf{s}),\dots,Z_J(\mathbf{s})$:\begin{equation}\mu_t(\mathbf{s})=\sum_{j=1}^J Z_j(\mathbf{s})\beta_{t,j},\end{equation} where $\beta_{t,j}$ is the time-varying effect of covariate $Z_j$ on day $t$. Potential covariates can be geographic variables such as elevation, spatiotemporal variables such as temperature, or polynomial functions of $\mathbf{s}$ and $t$ to model large-scale trends. The coefficient vectors $\mbox{\boldmath $\beta$}_t=[\beta_{t,1},\dots,\beta_{t,J}]^T\overset{iid}{\sim} \mbox{Normal}(\mbox{\boldmath $\beta$}_{0},\mbox{\boldmath $\Omega$})$ are modeled as random effects with mean $\mbox{\boldmath $\beta$}_0$ and covariance $\mbox{\boldmath $\Omega$}$ since each day has a different weather pattern. Conditioned on the mean, the errors $\mathbf{e}_t=[e_t(\mathbf{s}_1),\dots,e_t(\mathbf{s}_n)]^T$ are assumed to be independent $\mbox{Normal}(\vec{0},\mbox{\boldmath $\Sigma$}_t)$ with a spatial covariance matrix $\mbox{\boldmath $\Sigma$}_t$, that will depend on $\mu_t$ as described below. Note we are assuming conditional normality but not marginal normality, except for trivial cases where the covariance is not dependent on the mean. Furthermore we do not assume temporal covariances between the $\mbox{\boldmath $\beta$}_t$ covariates to focus on the spatial model, though inclusion would likely help estimation and forecasting the mean process.

The error process can have different variances and correlations across multiple spatial regimes. While there are many approaches to index the regimes (typically based on large-scale geographic features), we index the regimes by the mean process. Using the mean to determine the regime allows the covariances to vary smoothly over space and time in a simple but effective way. Let $\mbox{\boldmath $\Sigma$}_t$ be the spatial covariance matrix on day $t$, such that
\begin{equation}
\mathbb{C}\text{ov}[Y_t(\mathbf{s}_i),Y_t(\mathbf{s}_j)]=\mbox{\boldmath $\Sigma$}_{t,ij} = \tau_t^2(\mathbf{s}_i)\cdot\mathbb{I}(i=j)+\sigma_t(\mathbf{s}_i)\;\sigma_t(\mathbf{s}_j)\;\mathbf{R}_t^{NS}(\mathbf{s}_i,\mathbf{s}_j),\label{eq:covarfunc}
\end{equation}
where $\tau_t^2(\mathbf{s}_i)$ is the spatially-varying nugget effect, $\sigma^2_t(\mathbf{s}_i)$ is the spatially-varying partial sill, and $\mathbf{R}_t^{NS}(\mathbf{s}_i,\mathbf{s}_j)$ is a nonstationary spatial correlation function. 

For the nonstationary correlation matrix $\mathbf{R}_t^{NS}$, we use the method developed in \cite{paciorek2006} to construct a positive-definite, nonstationary correlation function using local kernel matrices. Define $\mathbf{C}_{t,i}$ to be a positive-definite local kernel matrix for the correlation between $\mathbf{s}_i$ and nearby locations.  For locations $\mathbf{s}_i$ and $\mathbf{s}_j$, their nonstationary correlation is determined by their local kernel matrices:
\begin{equation}
\mathbf{R}_t^{NS}(i,j) = |\mathbf{C}_{t,i}|^{1/4}|\mathbf{C}_{t,j}|^{1/4}\left|\frac{\mathbf{C}_{t,i}+\mathbf{C}_{t,j}}{2}\right|^{-1/2} \mathbf{R}^S(\sqrt{Q_{t,ij}})\label{eq:NonstatFunc}
\end{equation}
where $\mathbf{R}^S(\cdot)$ is a stationary correlation function of distance, and
\begin{equation}
Q_{t,ij}=(\mathbf{s}_i - \mathbf{s}_j)^T \left(\frac{\mathbf{C}_{t,i}+\mathbf{C}_{t,j}}{2}\right)^{-1}(\mathbf{s}_i - \mathbf{s}_j)
\end{equation}
is the nonstationary extension of the Mahalanobis distance. We assume that the local kernel matrix is isotropic,  so that $\mathbf{C}_{t,i}=\rho_t(\mathbf{s}_i) \mathbf{I}_2$ and
\begin{equation}
\mathbf{R}_t^{NS}(i,j)=\left\{\frac{4 \rho_t(\mathbf{s}_i) \rho_t(\mathbf{s}_j)}{[\rho_t(\mathbf{s}_i)+\rho_t(\mathbf{s}_j)]^2} \right\}^{p/4} \mathbf{R}^S\left\{ h_{ij}/\sqrt{\left[\rho_t(\mathbf{s}_i)+\rho_t(\mathbf{s}_j)\right]/2}\right\}
\end{equation} 
where $h_{ij} = ||\mathbf{s}_i - \mathbf{s}_j||$. Note that if the regimes are equivalent, $\mu_t(\mathbf{s}_i) = \mu_t(\mathbf{s}_j)$, then 
\begin{equation}
\mathbf{R}_t^{NS}(i,j) = \mathbf{R}^S\left(h_{ij} / \sqrt{\rho_t(\mathbf{s})} \right),
\end{equation} 
implying that $\rho_t(\cdot)$ determines the range for the correlation within a regime. The Mat\`{e}rn class of functions is a typical choice for $\mathbf{R}^S(r)$, of which we use the exponential function $\mathbf{R}^s(r)=\exp\{-r\}$ as a simple special case. 

For spatially-varying nugget, partial sill, and range, we elect to use mean-dependent link functions
\begin{equation}\tau_t^2(\mathbf{s}) = g_1[\mu_t(\mathbf{s});\mbox{\boldmath $\eta$}]; \; \sigma_t(\mathbf{s}) = g_2[\mu_t(\mathbf{s});\mbox{\boldmath $\eta$}]; \; \rho_t(\mathbf{s}) = g_3[\mu_t(\mathbf{s});\mbox{\boldmath $\eta$}] \label{eq:CovLinks} \end{equation}
that depend upon a set of unknown parameters $\mbox{\boldmath $\eta$}$. Other methods of obtaining nonstationarity would be to estimate parameters through the covariates \citep{risser2015}, but we present here a flexible ableit highly specialized model where the mean is the random spatial process varying across days to determine the degree of nonstationarity. While any nonnegative link function would suffice, we elect to use an exponential link function. The link function does not need to be monotonic, and much more sophisticated link functions could be used. For the exposition and simulation study, we assume the link is linear in the mean, such that $g_i[\mu_t(\mathbf{s});\mbox{\boldmath $\eta$}] = \exp[a_i + b_i \mu_t(\mathbf{s})]$. For the data analysis, we also explore the link that is linear in the log-mean $g_i[\mu_t(\mathbf{s});\mbox{\boldmath $\eta$}] = \exp[a_i + b_i \log \mu_t(\mathbf{s})]$, which leads to a better fit. 

In both cases, there are six parameters $\mbox{\boldmath $\eta$}=(a_1,b_1,a_2,b_2,a_3,b_3)$ to be estimated. The $a_i$ parameters affect the covariance parameters where either at the baseline mean equal to zero or (if $b_i = 0$) under stationary assumptions. The $b_i$ parameters indicate how the mean affects the local covariance function, and because the mean is the only spatially varying component of the covariance function, the $b_i$ parameters control the degree of nonstationarity in the model. The $(a_1, b_1)$ parameters affect the nugget of the covariance function, the $(a_2,b_2)$ parameters affect the sill, and the $(a_3,b_3)$ parameters affect the range of correlated sites. However, the entire set of parameters should be considered when interpreting how the local covariance function changes with respect to the mean.

\section{Computation}\label{sec:computation}
In vector form, $\mathbf{Y}_t = \mbox{\boldmath $\mu$}_t + \mathbf{e}_t = \mathbf{Z} \mbox{\boldmath $\beta$}_t + \mathbf{e}_t,$ where the design matrix $\mathbf{Z}$ contains the covariates $\mathbf{Z}_{ij}=Z_j(\mathbf{s}_i)$. In our representaion, the covariances change over time only via $\mbox{\boldmath $\beta$}_t$, so we denote the covariance matrix $\mathbb{C}\text{ov}(\mathbf{e}_t|\mbox{\boldmath $\beta$}_t) = \mbox{\boldmath $\Sigma$}(\mbox{\boldmath $\beta$}_t,\mbox{\boldmath $\eta$})$. The penalized likelihood function is then
\begin{equation}
L(\mbox{\boldmath $\eta$};\mbox{\boldmath $\beta$}_1,\dots,\mbox{\boldmath $\beta$}_m) = \prod_{t=1}^m \phi[\mathbf{Y}_t|\mbox{\boldmath $\beta$}_t,\mbox{\boldmath $\Sigma$}(\mbox{\boldmath $\beta$}_t,\mbox{\boldmath $\eta$})] \cdot \phi(\mbox{\boldmath $\beta$}_t|\mbox{\boldmath $\beta$}_{0},\mbox{\boldmath $\Omega$})\label{eq:likelihood}
\end{equation}
where $\phi$ is the multivariate normal density function. While construction of this likelihood function is well-reasoned, we acknowledge that we cannot guarantee the likelihood is not multimodal as parameters are only weakly identifiable, particularly when the nonstationary spatial covariances are relatively small. As a result, we carefully restrict parameter estimation in this study through profile maximum likelihood estimation methods, as described below. Bayesian methods could be explored to address uncertainty in the parameter estimations, but maximum likelihood is chosen because of the computational costs.

\subsection{The Stationary Model}\label{sec:StatModel}
When $b_1=b_2=b_3=0$, the covariance function in Equation \eqref{eq:NonstatFunc} loses dependence on the mean and becomes stationary. For the stationary model, we have $\mbox{\boldmath $\eta$}\equiv\mbox{\boldmath $\eta$}_0 = (a_1,0,a_2,0,a_3,0)$ and $\mathbb{C}\text{ov}(\mathbf{Y}_t|\mbox{\boldmath $\beta$}_t) \equiv \mbox{\boldmath $\Sigma$}(\mbox{\boldmath $\eta$}_0)$ for all $t$. Furthermore, conditioned on estimated covariance parameters $\hat{\mbox{\boldmath $\eta$}}_0$, the maximum penalized likelihood estimate of $\mbox{\boldmath $\beta$}_t$ is 
\begin{equation} 
\hat{\mbox{\boldmath $\beta$}}_t(\hat{\mbox{\boldmath $\eta$}}_0)= \left(\mathbf{Z}^T\mbox{\boldmath $\Sigma$}(\hat{\mbox{\boldmath $\eta$}}_0)^{-1}\mathbf{Z} + \mbox{\boldmath $\Omega$}^{-1}\right)^{-1}\mathbf{Z}^T\mbox{\boldmath $\Sigma$}(\hat{\mbox{\boldmath $\eta$}}_0)^{-1}\mathbf{Y}_t. \label{eq:BetaEstim} 
\end{equation}
To estimate $a_1$, $a_2$, and $a_3$, we numerically maximize \eqref{eq:likelihood} with $b_1$, $b_2$, and $b_3$ set to 0 and iterate between updating $\mbox{\boldmath $\eta$}_0$ and $\mbox{\boldmath $\beta$}_t$ until the likelihood converges to its maximum.

\subsection{The Nonstationary Model}\label{sec:NonstModel}
For the full nonstationary model without the restriction that $b_1=b_2=b_3=0$, the optimization is challenging because $\mbox{\boldmath $\beta$}_t$ appears in the covariance and a closed-form solution like \eqref{eq:BetaEstim} is no longer available. However, the estimator in \eqref{eq:BetaEstim} provides a reasonable approximation and propose a computationally-efficient one-step process. The one-step approximation fixes $\mbox{\boldmath $\beta$}_t$ equal to the estimated $\hat{\mbox{\boldmath $\beta$}}_t$ from the stationary model in Section \ref{sec:StatModel} and optimizes \eqref{eq:likelihood} with respect to $\mbox{\boldmath $\eta$} = (a_1,b_1,a_2,b_2,a_3,b_3)$. We choose not to iterate between estimating $\mbox{\boldmath $\beta$}_t$ and $\mbox{\boldmath $\eta$}$, making this method a one-step approximation to the profile maximum likelihood estimate. 

We found that multiple iterations of refitting the mean given the previously estimated covariance matrix did not guarantee convergence in the estimates, tended to overfit the mean process, and negatively affected predictive accuracy. However, the ultimate goal is to compare the stationary and nonstationary models to test for significant differences. Both the stationary and the one-step approximation have the same mean process but different spatial covariance models that can be compared to determine if there exists significant nonstationarity. Conversely, multiple iterations would lead to different mean processes and covariances, so the increased degrees of freedom make testing for nonstationarity more complicated.

To obtain the full maximum likelihood estimates for the nonstationary model, we also numerically optimize the likelihood in \eqref{eq:likelihood} over the full $(\mbox{\boldmath $\eta$};\mbox{\boldmath $\beta$}_1,\dots,\mbox{\boldmath $\beta$}_m)$ parameter set. Numerically maximizing over all of the $\mbox{\boldmath $\beta$}_t$ coefficients is time-consuming, especially as the number of days increases, so we use the one-step approximation estimates as starting values to achieve convergence within fewer steps. In practice, this full maximization would be performed after deeming the one-step approximation is deemed significantly different from the stationary model.
\section{Evaluation of models}
Once models are fit to the data, we evaluate the nonstationary model to determine if it significantly improves upon the stationary model. First, we derive a hypothesis test to determine if the mean-dependent terms of the nonstationary model are significant. Then, we obtain the prediction distribution that uses the fitted model to estimate the response over the spatial domain.
\subsection{Test for Nonstationarity}\label{sec:NStest}
Under the restriction $b_1 = b_2 = b_3 = 0$, the spatial covariance no longer depends on the mean, leading to a stationary covariance model. We can hence test the hypotheses
\begin{eqnarray}
H_0: & & b_1 = b_2 = b_3 = 0,\nonumber\\
H_A: & & b_j \neq 0,\text{ for at least one }j \nonumber
\end{eqnarray}
using a likelihood ratio test. Let $\hat{\mbox{\boldmath $\eta$}}_0$ and $\hat{\mbox{\boldmath $\beta$}}_{t(0)}$ be the parameter estimates under stationarity, and $\hat{\mbox{\boldmath $\eta$}}_1$ and $\hat{\mbox{\boldmath $\beta$}}_{t(1)}$ be the parameter estimates under nonstationarity. The Wilks' test statistic $\chi^2 = 2 \cdot [\log L(\hat{\mbox{\boldmath $\eta$}}_1;\hat{\mbox{\boldmath $\beta$}}_{1(1)},\dots,\hat{\mbox{\boldmath $\beta$}}_{m(1)}) - \log L(\hat{\mbox{\boldmath $\eta$}}_0;\hat{\mbox{\boldmath $\beta$}}_{1(0)},\dots,\hat{\mbox{\boldmath $\beta$}}_{m(0)})]$ is proportional to the difference in the log-likelihood:
\begin{eqnarray}
\chi^2 = \sum_{t=1}^m\!\!\!\!& &\!\!\!\!\left\{\log|\mbox{\boldmath $\Sigma$}_{t(1)}|+\left[\mathbf{Y}_t-\mathbf{Z}\mbox{\boldmath $\beta$}_{t(1)}\right]^T \mbox{\boldmath $\Sigma$}_{t(1)}^{-1}\left[\mathbf{Y}_t-\mathbf{Z}\mbox{\boldmath $\beta$}_{t(1)}\right] + \mbox{\boldmath $\beta$}_{t(1)}^T\mbox{\boldmath $\Omega$}^{-1}\mbox{\boldmath $\beta$}_{t(1)}\right\} \nonumber \\
& &\!\!\!\!-\left\{\log|\mbox{\boldmath $\Sigma$}_{t(0)}| + \left[\mathbf{Y}_t-\mathbf{Z}\mbox{\boldmath $\beta$}_{t(0)}\right]^T \mbox{\boldmath $\Sigma$}_{t(0)}^{-1}\left[\mathbf{Y}_t-\mathbf{Z}\mbox{\boldmath $\beta$}_{t(0)}\right] + \mbox{\boldmath $\beta$}_{t(0)}^T\mbox{\boldmath $\Omega$}^{-1}\mbox{\boldmath $\beta$}_{t(0)}\right\}. \label{chisqStat}
\end{eqnarray}
Under $H_0$, the Wilks' test statistic \citep{wilks1938} should follow a Chi-squared distribution with degrees of freedom equal to the difference in dimensionality of the parameter space between the full and reduced model. Since stationarity occurs by restricting three parameters, the difference in dimensionality would be three, so we would reject the hypothesis that the data are stationary at the $\alpha$ significance level if the test statistic exceeds the $\alpha$-level critical value for the Chi-squared distribution with $df=3$ degrees of freedom. This method could then be generalized to determine the significance of a wide variety of nested models, depending on the defined link functions in \eqref{eq:CovLinks} and which parameters are restricted.

In theory, the test statistic converges in distribution to the Chi-squared with $df=3$ degrees of freedom as the number of data points approaches infinity. In practice, the convergence rate may be too slow to assume the test statistic follows the Chi-squared. Furthermore, the inherent connection between the covariance and the mean indicate that the computation method from Section \ref{label:sec:computation} could affect the degrees of freedom of the test statistic. Under the one-step approximation, $\mbox{\boldmath $\Sigma$}_{t(0)}$ and $\mbox{\boldmath $\Sigma$}_{t(1)}$ would be the same, but under the full MLE computation, they would be different. Hence, the test statistic would most closely follow the Chi-squared distribution with $df=3$ degrees of freedom when using the one-step approximation method to estimate the covariance parameters. In Section \ref{sec:NStestSim}, we explore the convergence of the Type I error and power for this test statistic using both computation methods and increasing sample sizes.

\subsection{Prediction Distribution} \label{sec:predDist}
Let $\mathbf{Y}_0$ be the unknown values of the responses at a set of prediction sites, and $\mathbf{Y}_1$ be the known values of the responses at a set of training sites for a given day. Following the general format of spatial kriging, the training and prediction sites are stacked into a joint multivariate normal distribution:
\begin{equation}
\begin{pmatrix} \mathbf{Y}_0 \\ \mathbf{Y}_1 \end{pmatrix} \sim \mbox{Normal}\left[\begin{pmatrix} \mbox{\boldmath $\mu$}_0 \\ \mbox{\boldmath $\mu$}_1 \end{pmatrix}, \begin{pmatrix} \mbox{\boldmath $\Sigma$}_{00} & \mbox{\boldmath $\Sigma$}_{01} \\ \mbox{\boldmath $\Sigma$}_{10} &\mbox{\boldmath $\Sigma$}_{11} \end{pmatrix} \right]
\end{equation}
where $\mbox{\boldmath $\mu$}_0 = \mathbf{Z}_0 \mbox{\boldmath $\beta$}$ and $\mbox{\boldmath $\Sigma$}_{00} = \mbox{\boldmath $\Sigma$}(\mbox{\boldmath $\mu$}_0,\mbox{\boldmath $\eta$})$ are the marginal mean and covariance matrix for the prediction sites, $\mbox{\boldmath $\mu$}_1 = \mathbf{Z}_1 \mbox{\boldmath $\beta$}$ and $\mbox{\boldmath $\Sigma$}_{11} = \mbox{\boldmath $\Sigma$}(\mbox{\boldmath $\mu$}_1,\mbox{\boldmath $\eta$})$ are the marginal mean and covariance for the training sites, and $\mbox{\boldmath $\Sigma$}_{01}$ and $\mbox{\boldmath $\Sigma$}_{10}$ are the cross-covariance matrices between the prediction and training sites.

After the training data is used to estimate the mean function coefficients $\hat{\mbox{\boldmath $\beta$}}$ and covariance parameters $\hat{\mbox{\boldmath $\eta$}}$, we calculate the predicted responses $\hat{\mathbf{Y}}_0$ and their standard errors from the mean and variances in the conditional distribution of the prediction responses given the training responses:
\begin{equation}
(\mathbf{Y}_0|\mathbf{Y}_1,\hat{\mbox{\boldmath $\beta$}},\hat{\mbox{\boldmath $\eta$}}) \sim \mbox{Normal}\left(\mbox{\boldmath $\mu$}_0 + \mbox{\boldmath $\Sigma$}_{01} \mbox{\boldmath $\Sigma$}_{11}^{-1} (\mathbf{Y}_1 - \mbox{\boldmath $\mu$}_1), \mbox{\boldmath $\Sigma$}_{00} - \mbox{\boldmath $\Sigma$}_{01} \mbox{\boldmath $\Sigma$}_{11}^{-1} \mbox{\boldmath $\Sigma$}_{10}\right),\label{predictionDist}
\end{equation}
In \eqref{predictionDist}, the prediction distribution  is evaluated on a given day $t$, yet the notation has been suppressed for brevity. The prediction mean in \eqref{predictionDist} is our best prediction for the responses, with standard errors calculated from the square root of the diagonals of the covariance matrix. Note that training responses could be forecasted values using data from previous days at fixed sites, but the covariance matrix will need to incorporate temporal or autoregressive terms.

\section{Simulation study}\label{sec:SimStudy}
In this simulation study, we generate spatial data from the model, estimate parameters, and evaluate the predictive distribution. The purpose of the study is to compare the computational algorithms and understand the frequentist properties of the methods, with a type of stationary data as a baseline comparison. We show that we can estimate the covariance parameters accurately, detect nonstationarity, and predict more accurately than the stationary models.
\subsection{Simulated data generation}\label{sec:DataGen}
\begin{figure}[p]
\begin{center}
\includegraphics[height=0.4\textheight]{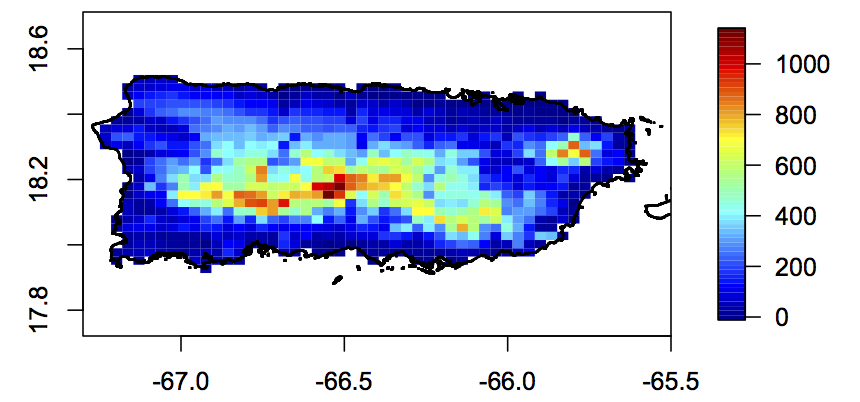}
\end{center}
\caption{Longitude, latitude, and elevation across the island of Puerto Rico. Longitude and latitudes given in coordinate degrees; elevation given in meters above sealevel. The northern and southern coasts of the island are coastal plains, separated by the mountainous Cordillero Central. Most storms approach Puerto Rico from the north and the east, so the northern and eastern coastal plains have typically higher precipitation than the southern and western coastal plains. The Sierra de Luquillo is in the northeastern part of the island, which contains Pico El Yunque and the El Yunque National Rainforest. The mountains experience higher precipitation due to orographic uplift of the moisture-laden trade winds, leading to extensive wet montane forests and a sharp precipitation gradient from the northeast to southwest.}
\label{fig:PRelev}
\end{figure}

The simulated data are split into a training set to fit the model and a testing set to evaluate predictions. The training data contains $n$ spatial locations with observations on $m$ days. The testing set has 100 randomly-selected locations for each day. All training and testing locations are uniformly distributed over a rectangular domain with longitudes in the range $(-67.3,-65.7)$ and latitudes in the range $(17.9,18.5)$, which is approximately the geographic area of Puerto Rico used in Section \ref{sec:DataAnalysis} (see Figure \ref{fig:PRelev}). We use $J=3$ covariates for the mean, using an intercept and spatial coordinates $\mathbf{s} = (s_1,s_2)^T$. When generating data, the coefficients $\mbox{\boldmath $\beta$}_t=(\beta_{t1},\beta_{t2},\beta_{t3})$ are fixed such that the intercept is constant, $\beta_{t1}=1$, and the spatial effects increase over time, $\beta_{t2}=\beta_{t3} = 2 \frac{t-1}{m-1}$. 

\begin{figure}[t]
\begin{center}
Stationary simulated data:\\
\includegraphics[width=0.45\textwidth]{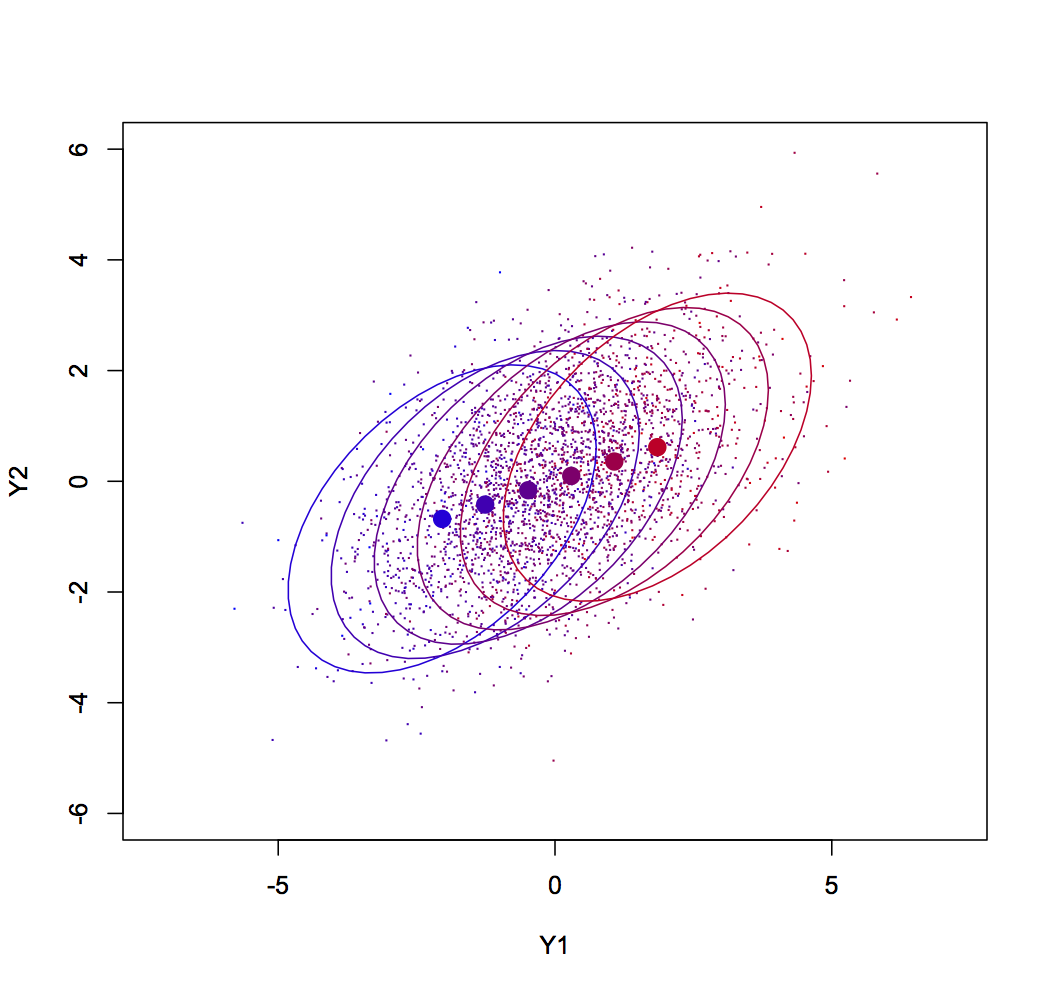}
\includegraphics[width=0.45\textwidth]{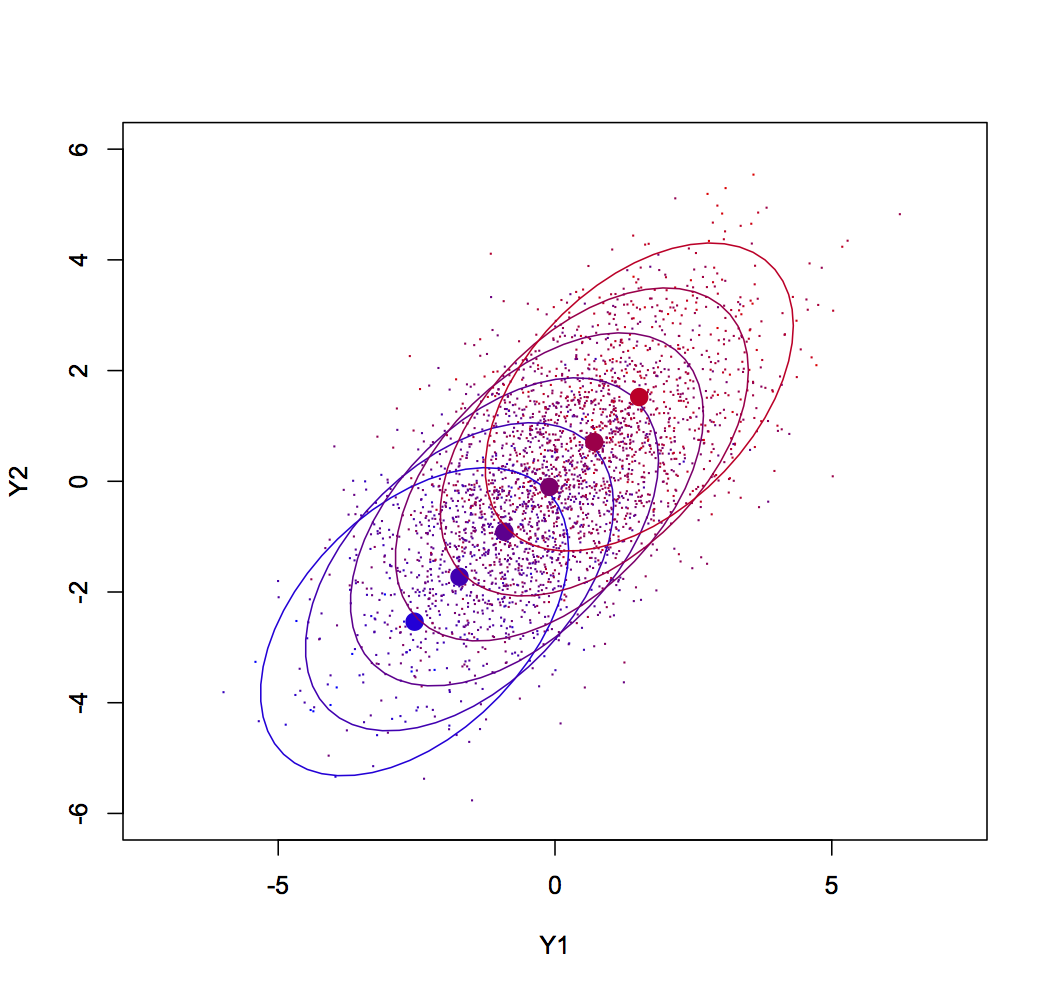}\\
Nonstationary simulated data:\\
\includegraphics[width=0.45\textwidth]{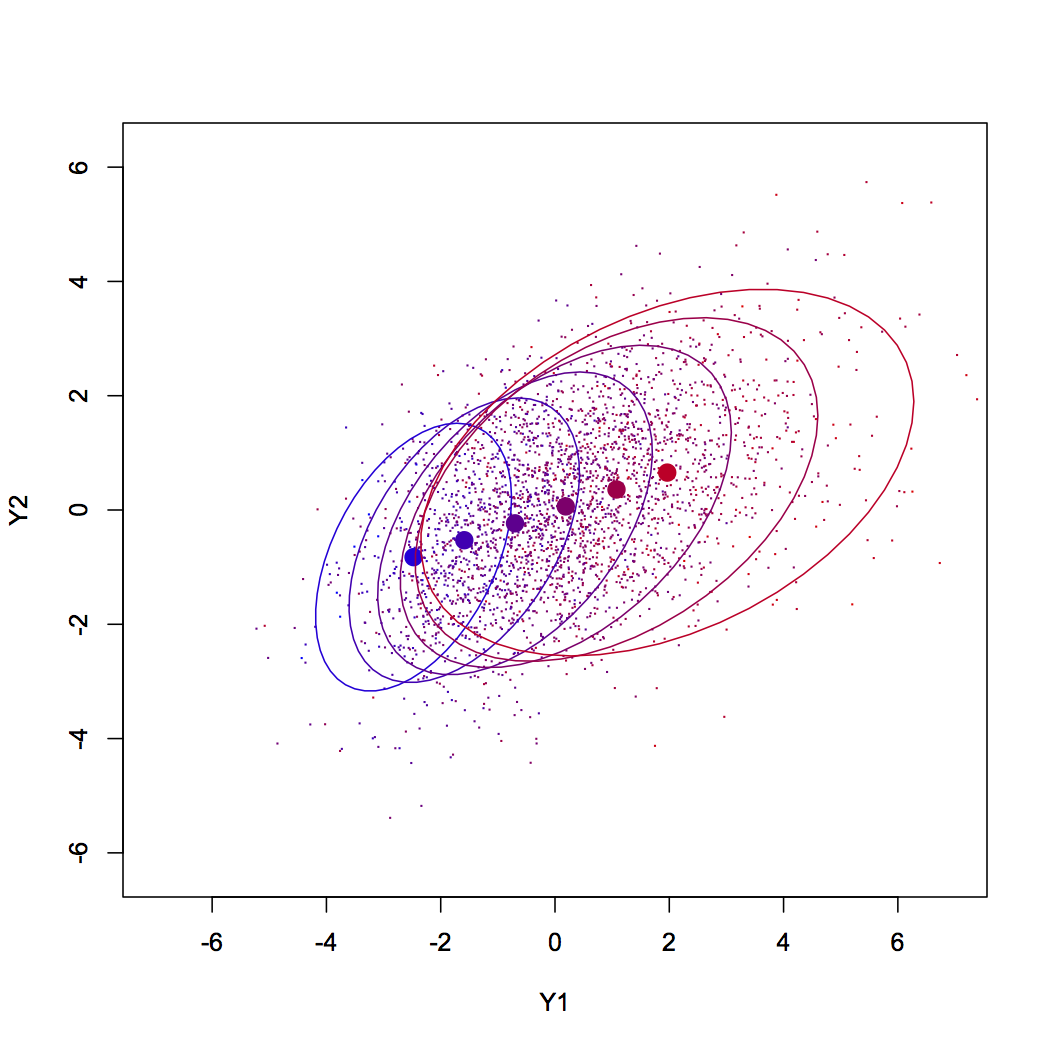}
\includegraphics[width=0.45\textwidth]{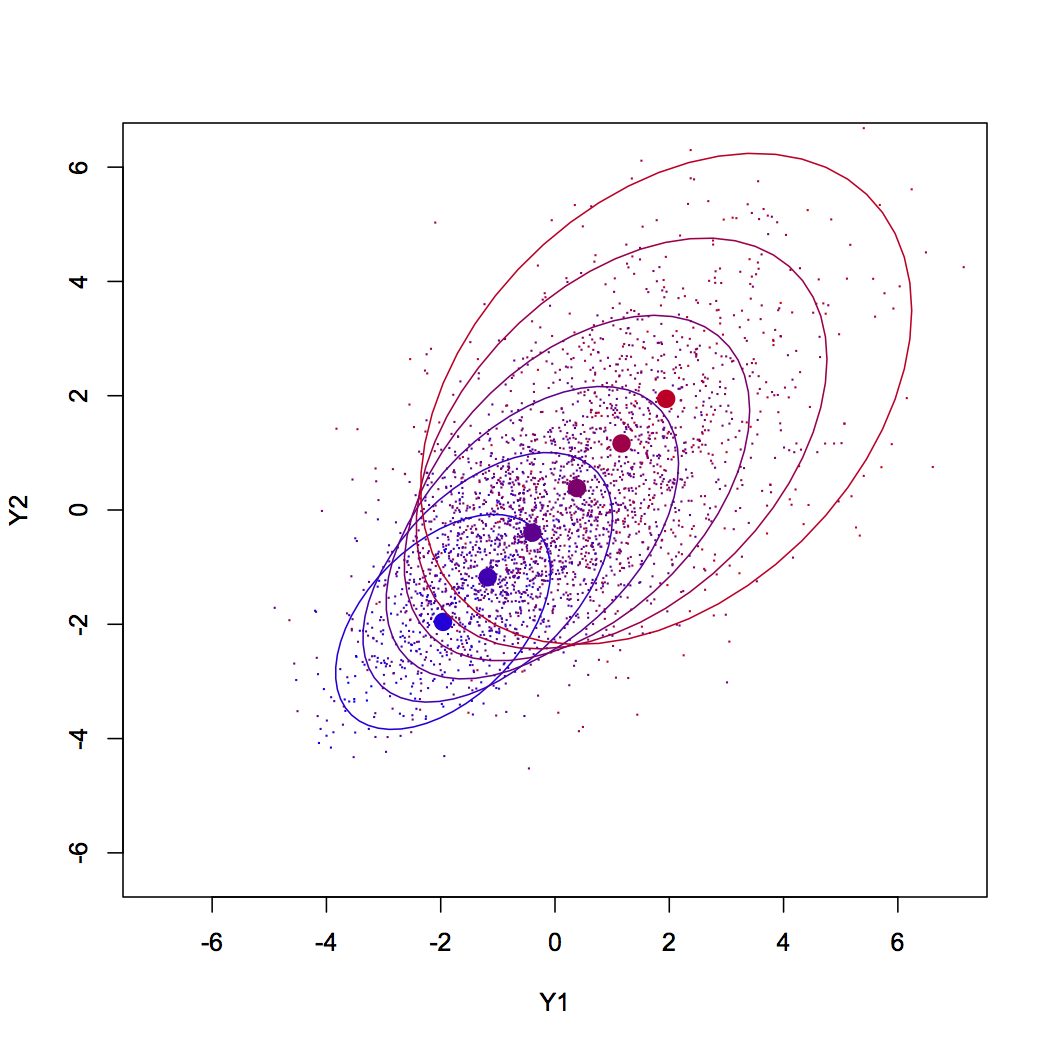}\\
\caption{Simulated data for two sites, generated as described in Section \ref{sec:DataGen}. Data in the right figures assume the sites are considered to be in the same regime ($\mu_1=\mu_2$). Data in the left figures assumes the sites are in different regimes ($\mu_1\neq\mu_2$). Top row figures used stationary covariance ($b_1=b_2=b_3=0$), and bottom row figures used mean-dependent nonstationary covariance. Colors indicate the magnitude of the mean, ranging from blue to red. The rings represent the central 95\% ellipse for the bivariate normal distribution with mean indicated by the large dot.  The smaller dots represent simulated responses, colored according to their mean. Nonstationarity is apparent in the changing size and shape of the rings.}
\label{fig:SimData}
\end{center}
\end{figure}

Given the $\mbox{\boldmath $\beta$}_t$ values for each day, the responses are generated using the covariance function in \eqref{eq:covarfunc}. The mean-independent covariance parameters remain fixed at $a_1 = -1.0$, $a_2 = 0.5$, and $a_3 = 4.0$. For nonstationary data, we set the mean-dependent parameters to be $b_1=0.1$, $b_2 = 0.5$, and $b_3 = -0.5$; for stationary data, these parameters are set to zero. Figure \ref{fig:SimData} illustrates how the simulations generate stationarity and nonstationarity under the chosen parameters. We consider both stationary and nonstationary data sets with $n=$ 50, 100 or 200 locations and $m=$ 5 or 10 days. For each set of experimental conditions, we create 200 replicates of the data simulations.

The parameters for the stationary and nonstationary models are estimated using the methods described in Section \ref{sec:computation}.  Since there are only $J=3$ parameters and the true $\mbox{\boldmath $\beta$}_t$ values are fixed for each simulation, we set $\mbox{\boldmath $\beta$}_0=\vec{0}$ and $\mbox{\boldmath $\Omega$}=e^{10}\mathbf{I}_3$ to effectively remove the random effects from the penalized likelihood function.

\subsection{Type I error and power rate for nonstationarity test} \label{sec:NStestSim}
We explore how effectively the test described in Section \ref{sec:NStest} determines a data set is nonstationary. Data sets are generated for a fixed $m=5$ days, with a varying number of spatial locations $n$. For each data set, we fit the data generated from the stationary model (i.e. $b_1 = b_2 = b_3 = 0$) and use the Chi-squared test statistic in \eqref{chisqStat} to test for non-stationarity at the 5\% significance level. The number of days is fixed at $m=5$, and the number of spatial locations $n$ is allowed to vary.  

\begin{figure}[p]
\begin{center}
\includegraphics[height=0.45\textheight]{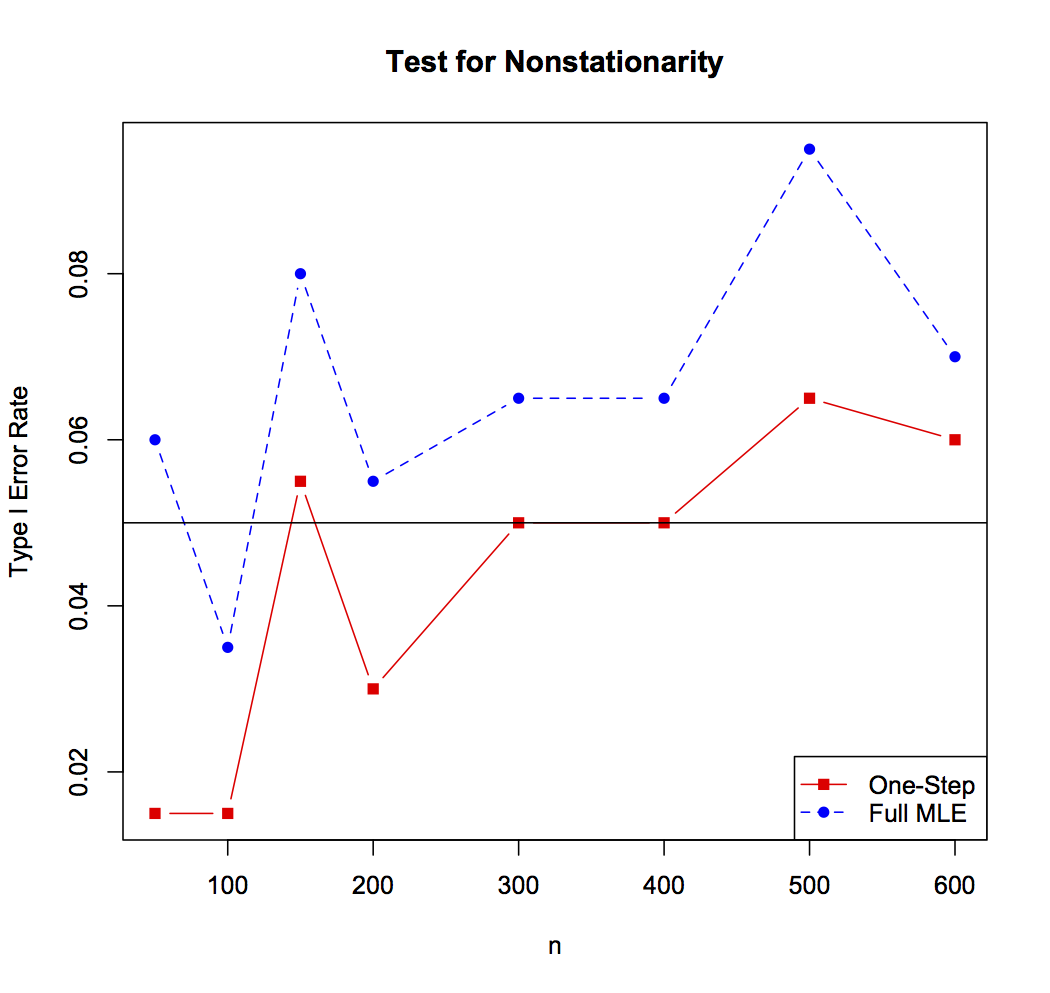}
\end{center}
(a) Type I error rate, against varying spatial locations $n$\\
\begin{center}
\includegraphics[height=0.45\textheight]{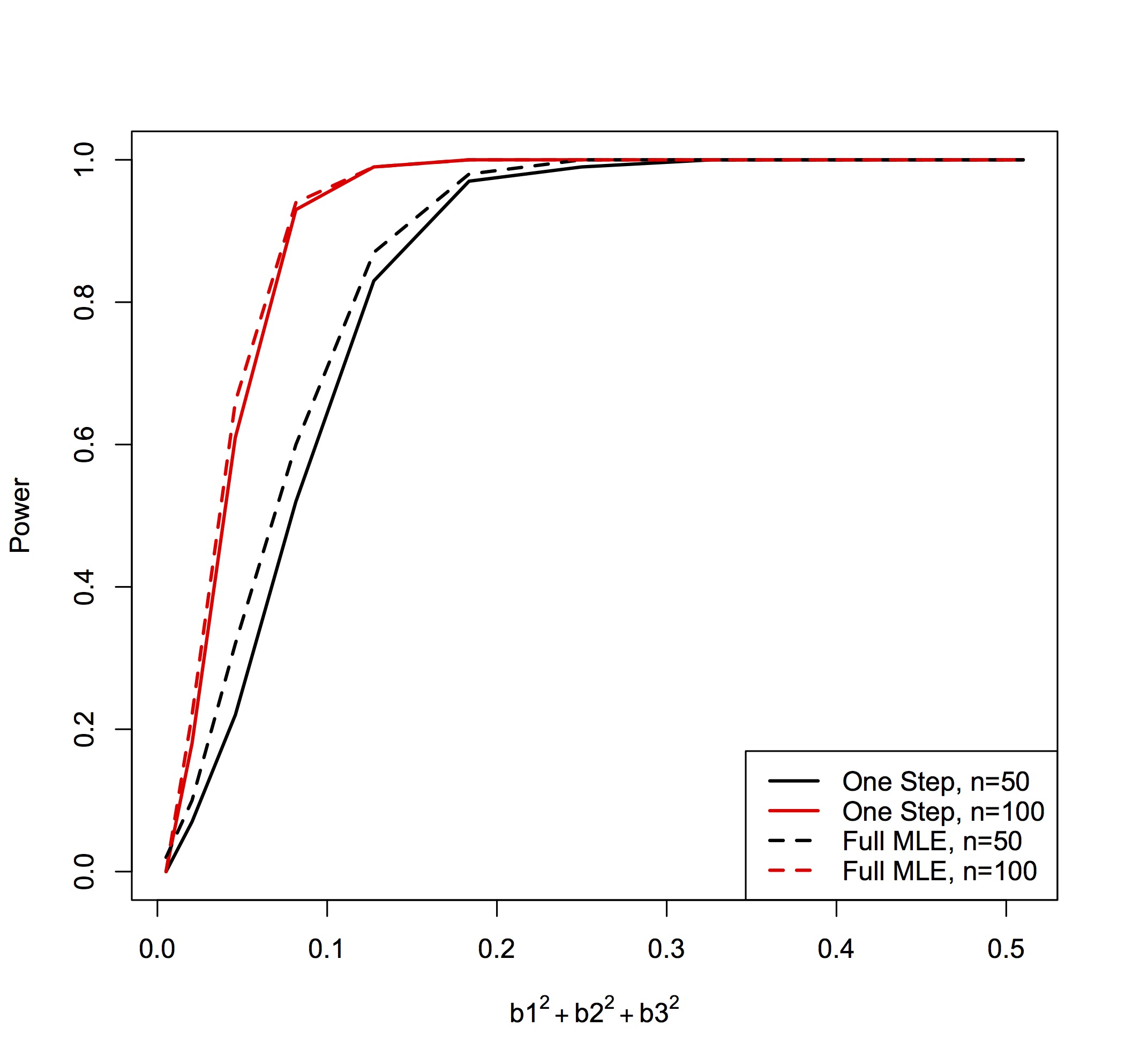}
\end{center}
(b) Power rate, against size of $b_1^2+b_2^2+b_3^2$\\
\caption{Type I error rate and power rate for nonstationarity tests at 5\% significance level. Each simulation uses $m=5$ days.}
\label{fig:TestError}
\end{figure}

To evaluate Type I error in Figure \ref{fig:TestError}(a), data is generated from a stationary model, with $b_1 = b_2 = b_3 = 0$. The one-step approximation converges to the expected 5\% error rate quicker than the full MLE. The full MLE generally has a larger Type I error rate than the expected 5\%, which indicates that the test will favor using the nonstationary model even when the data are stationary. 

To evaluate power in Figure \ref{fig:TestError}(b), data is generated from a range of nonstationarity, with $b_1=0.1*c$, $b_2=0.5*c$, and $b_3=-0.5*c$ and $c$ varied from 0.1 to 1.0. The power increases as the number of spatial locations increases from 50 to 100, and the one-step approximation typically has lower power than the full MLE.

\subsection{Parameter estimation accuracy} \label{sec:SimEtaMSE}
Table \ref{tbl:EtaEst} contains the mean squared error (MSE) for the parameter estimates summed over all of the covariance parameters; the MSE for each individual $\mbox{\boldmath $\eta$}$ parameter appears in Appendix \ref{app:etaMSEs}. As expected, for the stationary data, the stationary model is the most accurate, and for the nonstationary data, the nonstationary models are most accurate. The full MLE is the most accurate for all nonstationary cases, except for the smallest sample size. Table \ref{tbl:Times} contains the average times needed to fit the model. Between these tables we can see that the one-step approximation is faster than the full MLE while still being competitively accurate.

\subsection{Prediction distribution accuracy}
To measure prediction accuracy, we use prediction mean squared error, averaged over 100 prediction locations. The MSE is computed as the average squared difference between the testing data at the prediction locations (generated concurrently with the training data, as described in Section \ref{sec:DataGen}) and the predicted mean in \eqref{predictionDist} obtained from each model for each simulation. We calculate the percent improvement in MSE between each nonstationary model and the stationary model
\begin{equation}
\text{\% Improvement in MSE} = 100 \cdot \frac{MSE(\text{Stationary}) - MSE(\text{Nonstationary})}{MSE(\text{Stationary})} 
\end{equation}
averaged over the simulations. In addition to MSE for point prediction, we evaluate the deviance of the test data from the negative log-likelihood of the prediction distribution and calculate the differences between each nonstationary model and the stationary model, averaged over all of the simulations. For both measures, positive differences indicate that the nonstationary model yielded better predictions of the test data.

Table \ref{tbl:PredMSEScore} contains the percent improvement in MSE and log-likelihood for the simulated data. We observe that the nonstationary model fitted using full MLE is the best predictive model, but the one-step approximation is competitive. For the stationary data, the predictions from the nonstationary models are only slightly less accurate than the simpler stationary model (most likely due to overfitting to the training data), but the accuracy of the nonstationary data becomes comparable as the sample size increases. For the nonstationary data, the maximum likelihood estimates for the model have the most accurate predictive means and best overall fitting scores, but the one-step approximation is comparable.

\section{Data application: Precipitation data in Puerto Rico} \label{sec:DataAnalysis}
\begin{figure}[p]
\centering
\includegraphics[width=0.45\textwidth]{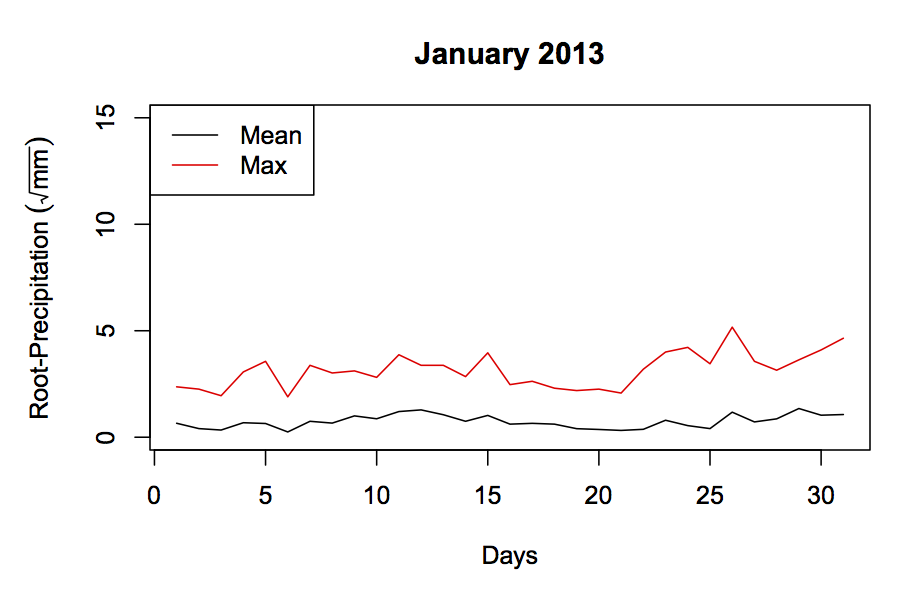}\hfill
\includegraphics[width=0.45\textwidth]{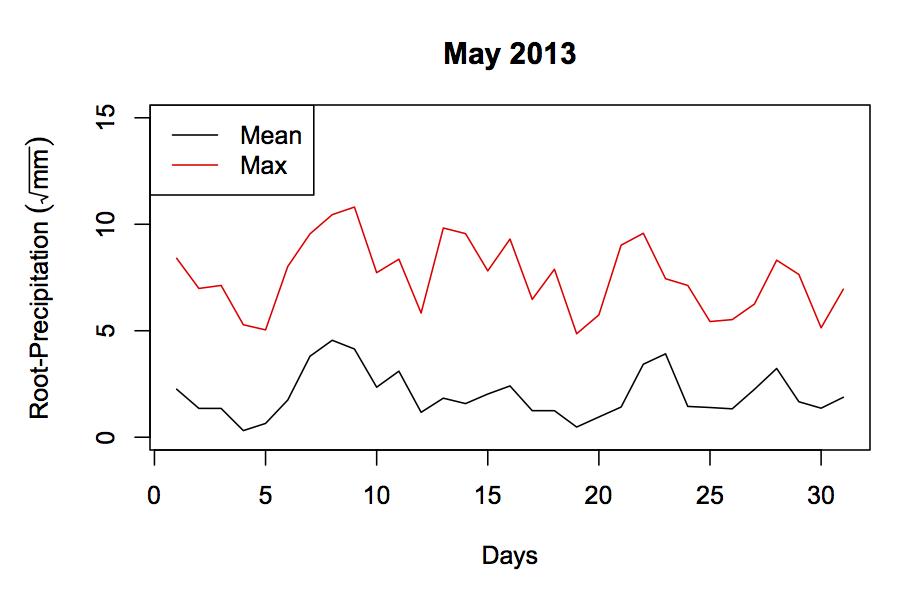}
\caption{Daily mean and max values for Puerto Rico data. Rain gauge measurements (in millimeters) obtained from weather stations in Puerto Rico over January 2013 (left) and over May 2013 (right). Data has been square-rooted.}
\label{fig:dry_rainy_seasons}
\end{figure}

We apply the mean-dependent nonstationary model to daily rain gauge measurements (in millimeters) from the island of Puerto Rico. The dataset comes from the Global Historical Climatology Network-Daily database \citep{menne2012}. We use the square-root of the measurements as the response throughout the data analysis. We consider two time periods: one during the dry season (January 2013) and one during the rainy season (May 2013). Figure \ref{fig:dry_rainy_seasons} depicts the daily mean and maximum values for the precipitation measurements over $n=47$ reporting weather stations. The rainy season has consistently higher means than the dry season. These two time periods were selected to illustrate the seasonal differences in Puerto Rico's climate, so the resulting best models may be different for other months or years.

\begin{figure}[p]
\begin{center}
\includegraphics[height=0.5\textwidth]{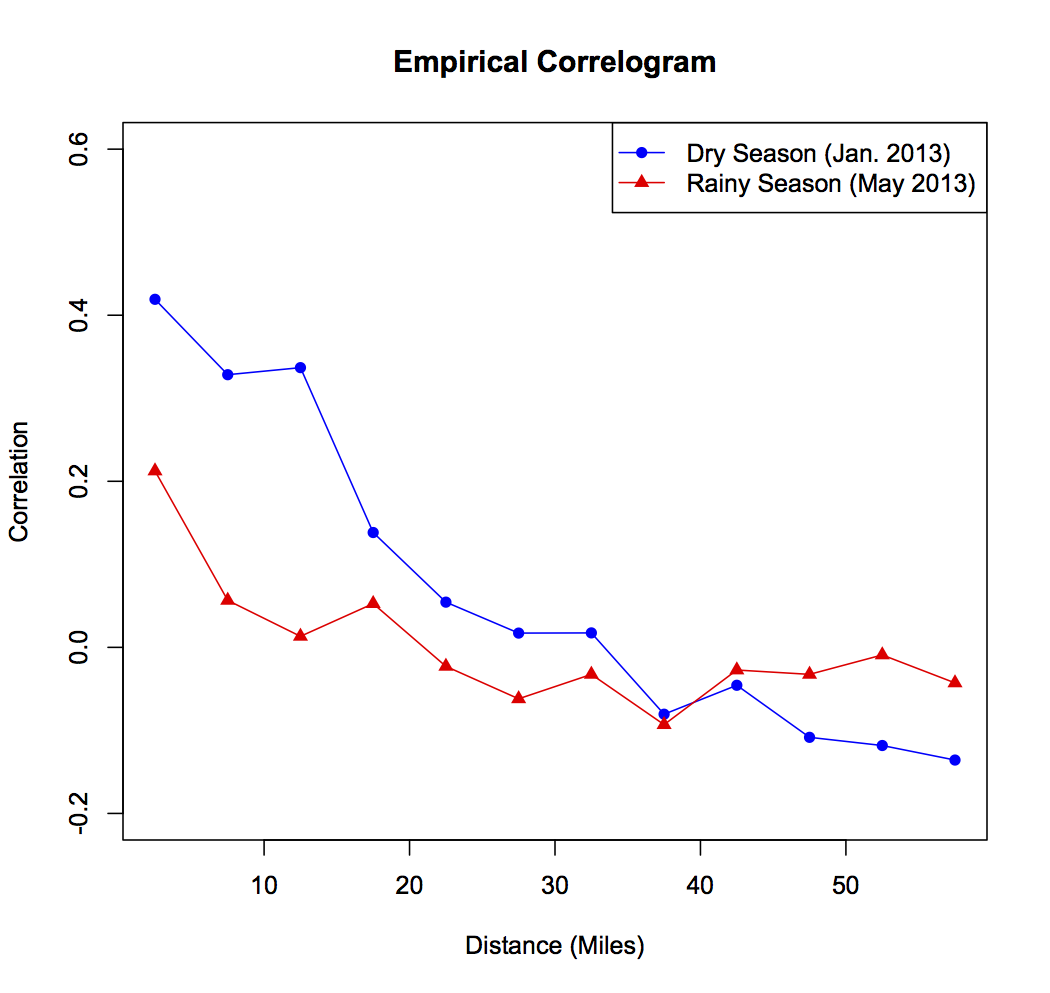}
\end{center}
\caption{Empirical correlograms for January 2013 and May 2013.}
\label{fig:dry_rainy_corr}
\end{figure}

\begin{figure}[p]
\begin{center}
\includegraphics[height=0.3\textheight]{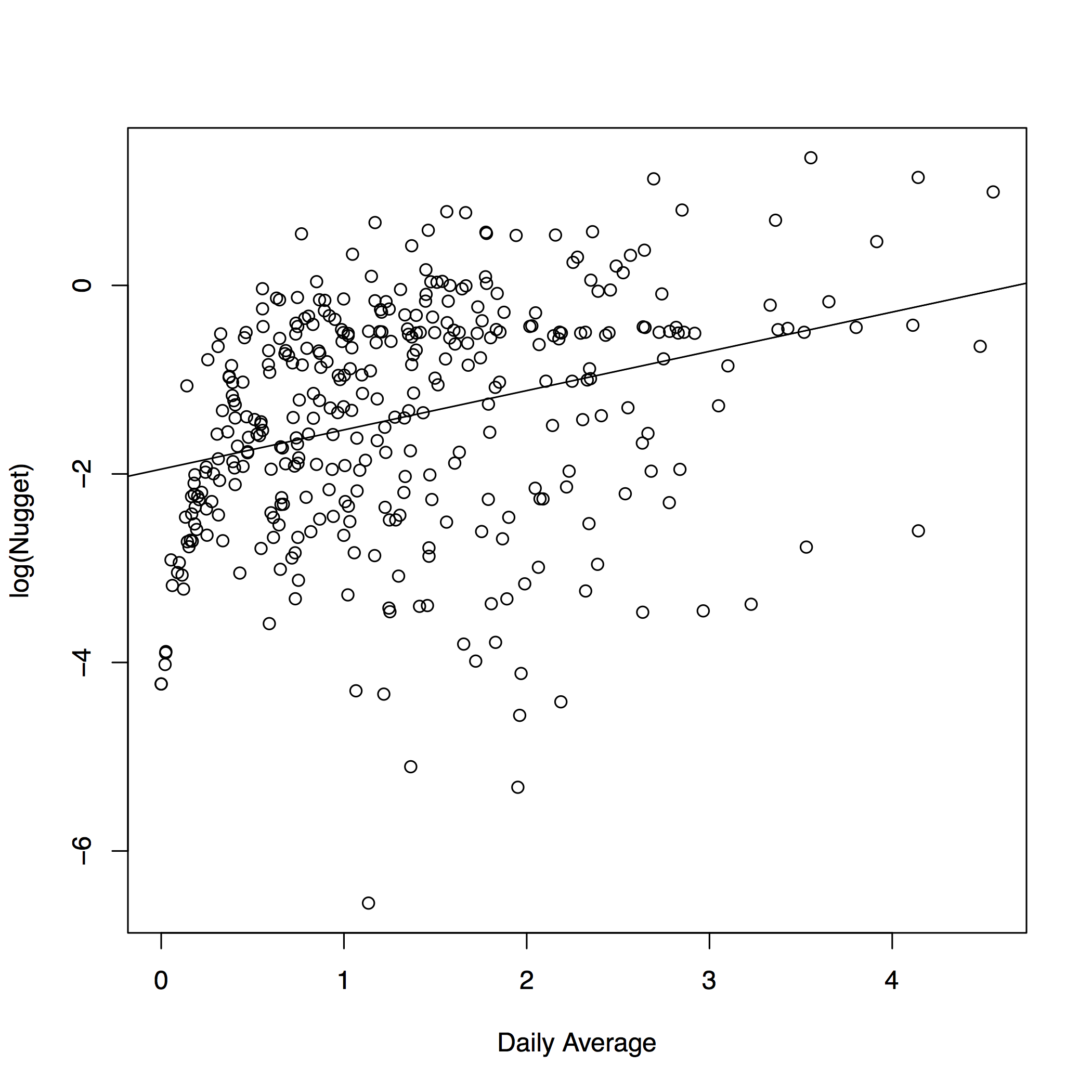}\hfill Nugget \hfill
\includegraphics[height=0.3\textheight]{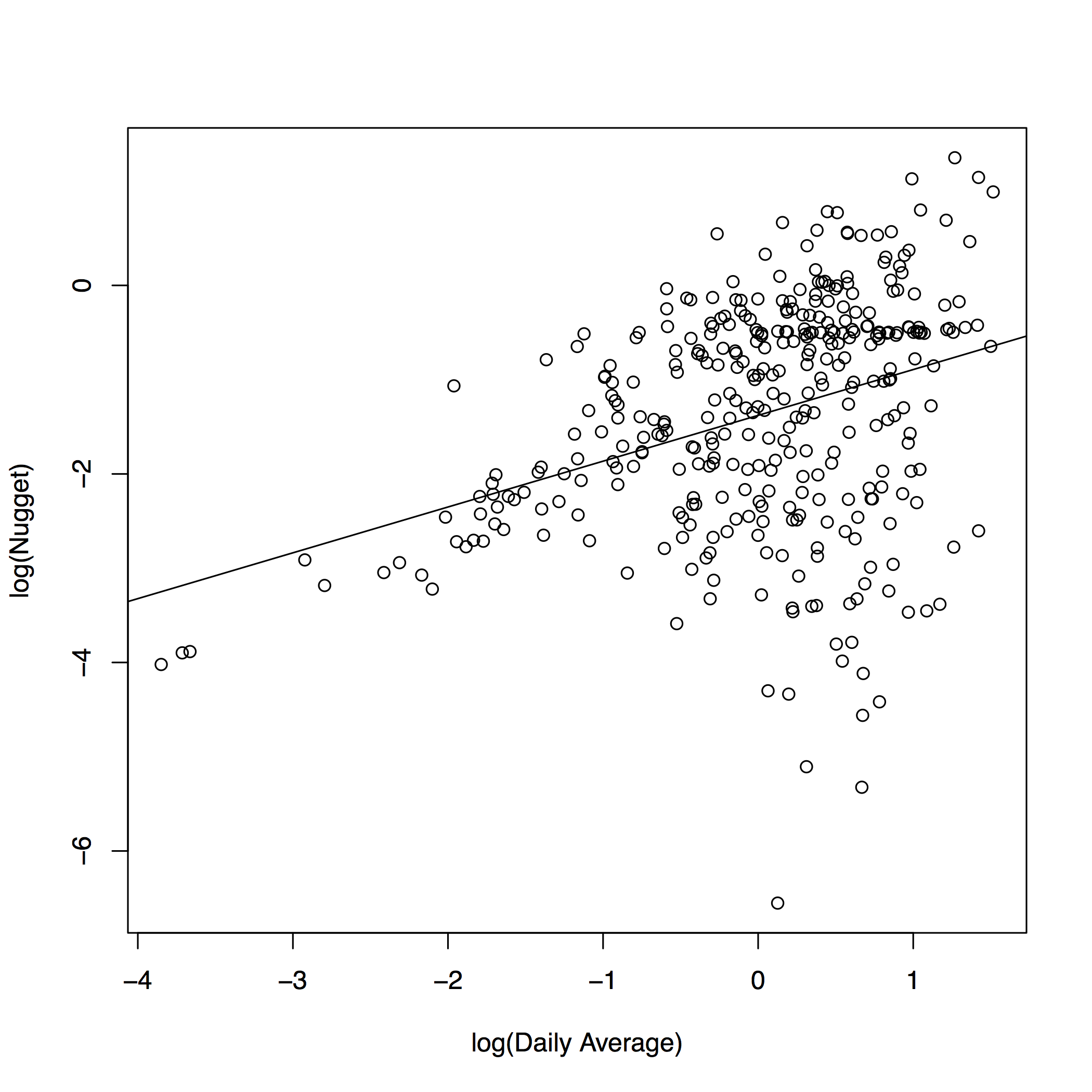}\\
\includegraphics[height=0.3\textheight]{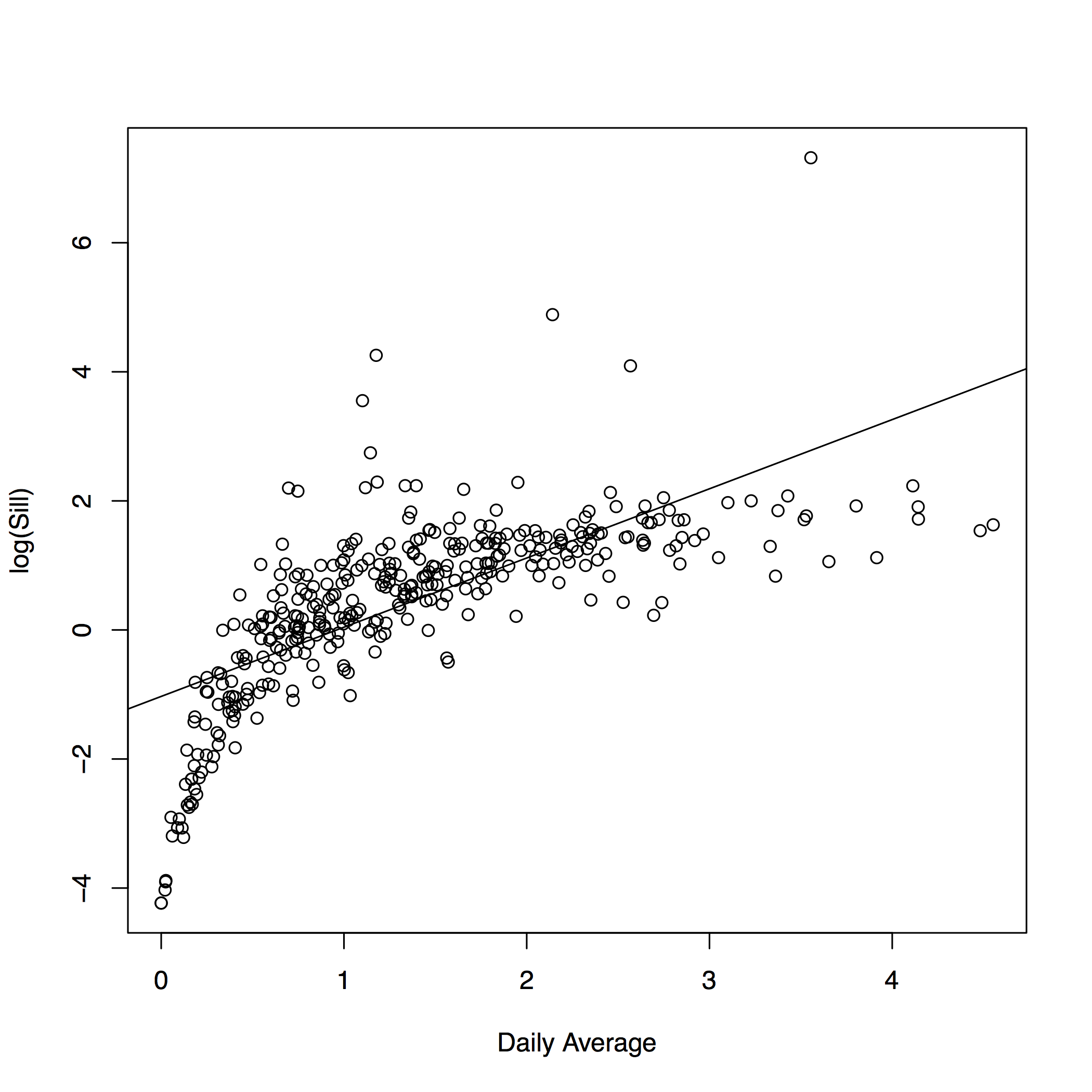}\hfill Sill \hfill
\includegraphics[height=0.3\textheight]{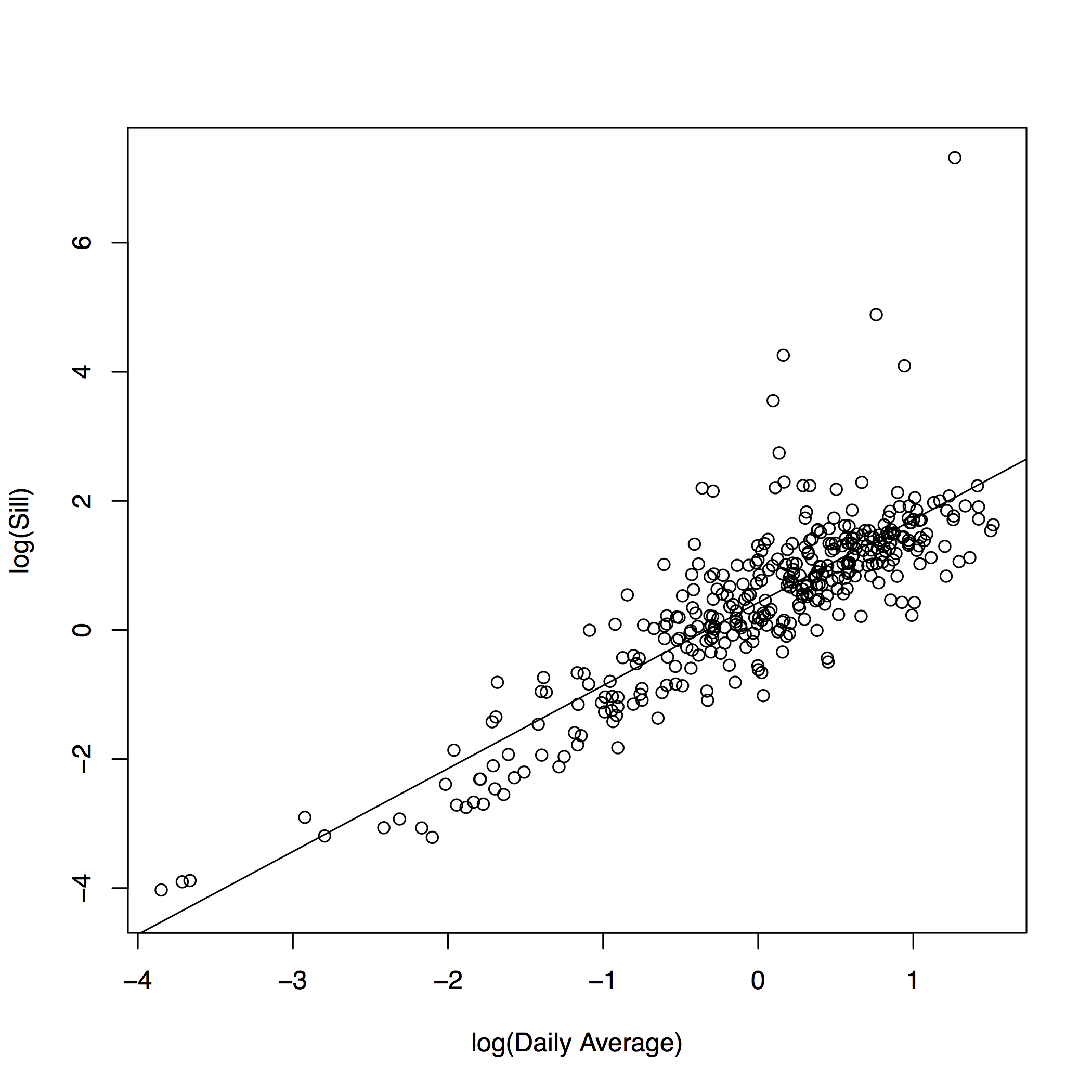}\\
\includegraphics[height=0.3\textheight]{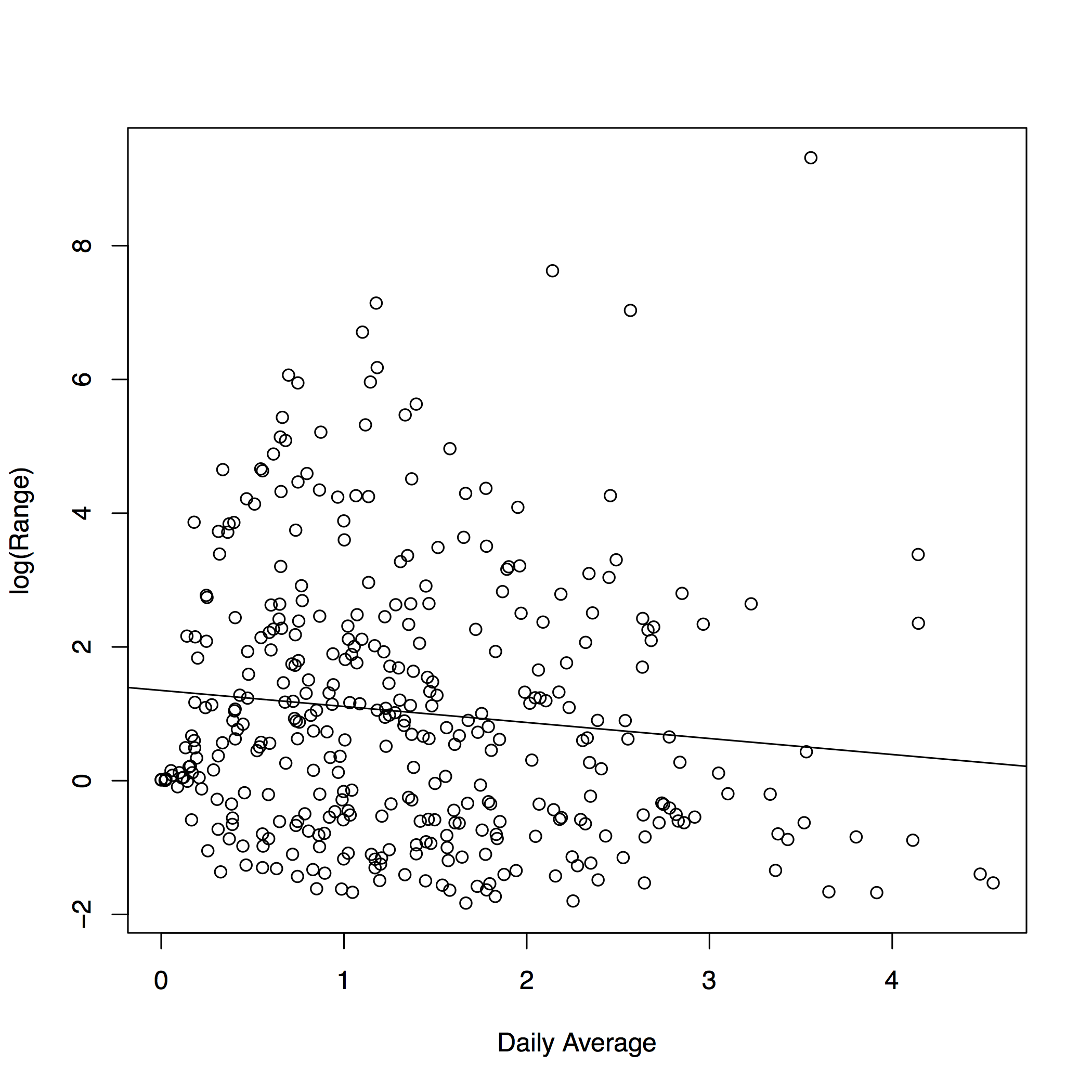}\hfill Range \hfill
\includegraphics[height=0.3\textheight]{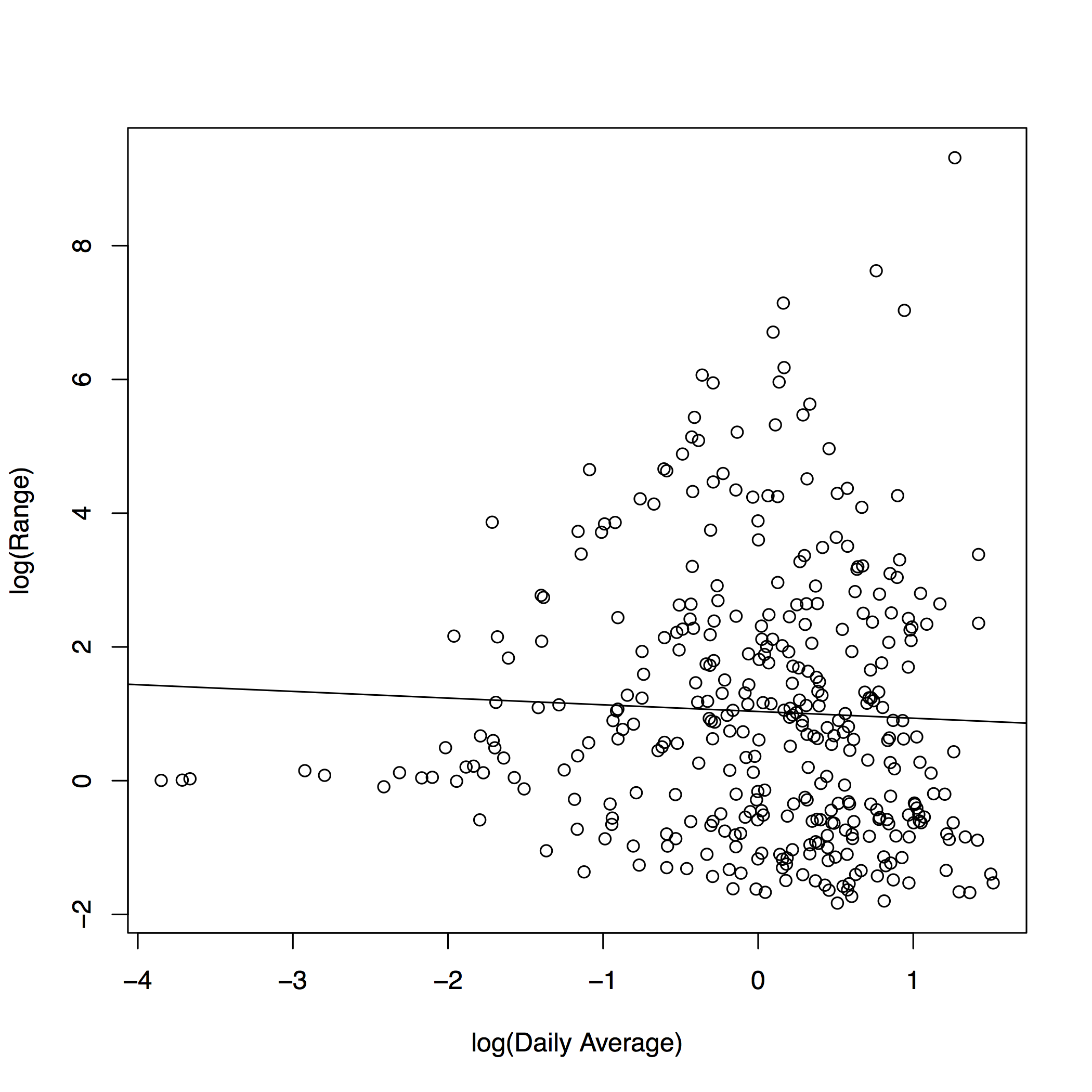}
\end{center}
\caption{Daily estimates of the stationary exponential covariance function parameters. Fitted for each day in 2013, and log-transforms of the parameter estimates are plotted against the daily average or the log-transform of the daily average.}
\label{fig: PR_nsr}
\end{figure}

Figure \ref{fig:dry_rainy_corr} shows significant differences in empirical correlograms, so we will fit the mean-dependent nonstationary model separately for each month. For the covariance parameters in \eqref{eq:CovLinks}, we perform an exploratory analysis to determine the proper link functions. For each day in 2013, we estimated a simple stationary model with exponential covariance function to obtain daily estimates for the nugget, sill, and range. These estimates are plotted against the daily averages in Figure \ref{fig: PR_nsr}. The log-transformations on nugget and sill indicate that exponential link functions should be used, $g_i(\mu) = \exp(a_i + b_i \mu)$, which we will refer to as the mean-dependent nonstationary (MDNS) fitted model in the analysis. However, log-transformations on the average lead to better linear relationships with the nugget and sill, so we will also consider $g_i(\mu) = \exp[a_i + b_i \log(1+\mu)]$, which we will refer to as the log-mean-dependent nonstationary (L-MDNS) fitted model. For the range, the log-transformation indicates that MDNS fit would work, but the relationship to the mean may be too weak. As a result, we will also consider fitting the model where $b_3=0$ is fixed, which would make the range stationary.

To fit the regime process, we use the longitude $s_1$, latitude $s_2$, and elevation $e(\mathbf{s})$ (see Figure \ref{fig:PRelev}). The mean is then regressed against either $p=4$ linear predictors $Z(\mathbf{s}) = [1, s_1, s_2, e(\mathbf{s})]^T$ or $p=7$ quadratic predictors $Z(\mathbf{s}) = [1, s_1, s_2, e(\mathbf{s}), s_1^2, s_1s_2, s_2^2]^T$. For the random-effects distribution, we compute the estimated $\mbox{\boldmath $\beta$}$ for each day and set $\mbox{\boldmath $\beta$}_0$ and $\mbox{\boldmath $\Omega$}$ to the sample mean and covariance of the estimates. We did not consider temporal effects mainly because there were no clear temporal trends, outside of the seasonal differences observed in Figure \ref{fig:dry_rainy_seasons}. 

We used five-fold cross-validation to evaluate the methods. For each fold, a random selection of locations are removed as testing data. The model is fit using the remaining sites as the training data. The fitted model is then used to construct the predictive distribution at the test sites. We evaluate the accuracy of the predictive distributions using three measures: the average log-likelihood score of the prediction distribution evaluates overall fit of the distribution, the mean squared error MSE measures the difference between the observed values and prediction mean, and the percent coverage of the observed values using 95\% probability prediction intervals. We also calculate the 5\%, 50\%, and 95\% quantiles of the standard errors to illustrate how the nonstationary models compare to the stationary models in terms of precision.

Negative values can be a potential issue with the raw precipitation data analysis. Our model may predict negative values, which is not physically possible. This issue could be remedied using a zero-truncated normal distribution \citep{stein1992, militino1999}, but this distribution is too computationally inefficient to use on such a large dataset. For practical purposes, we simply threshold negative predictions to zeroes. 

\subsection{January 2013 data}
\begin{figure}[p]
\begin{center}
\includegraphics[height=0.3\textheight]{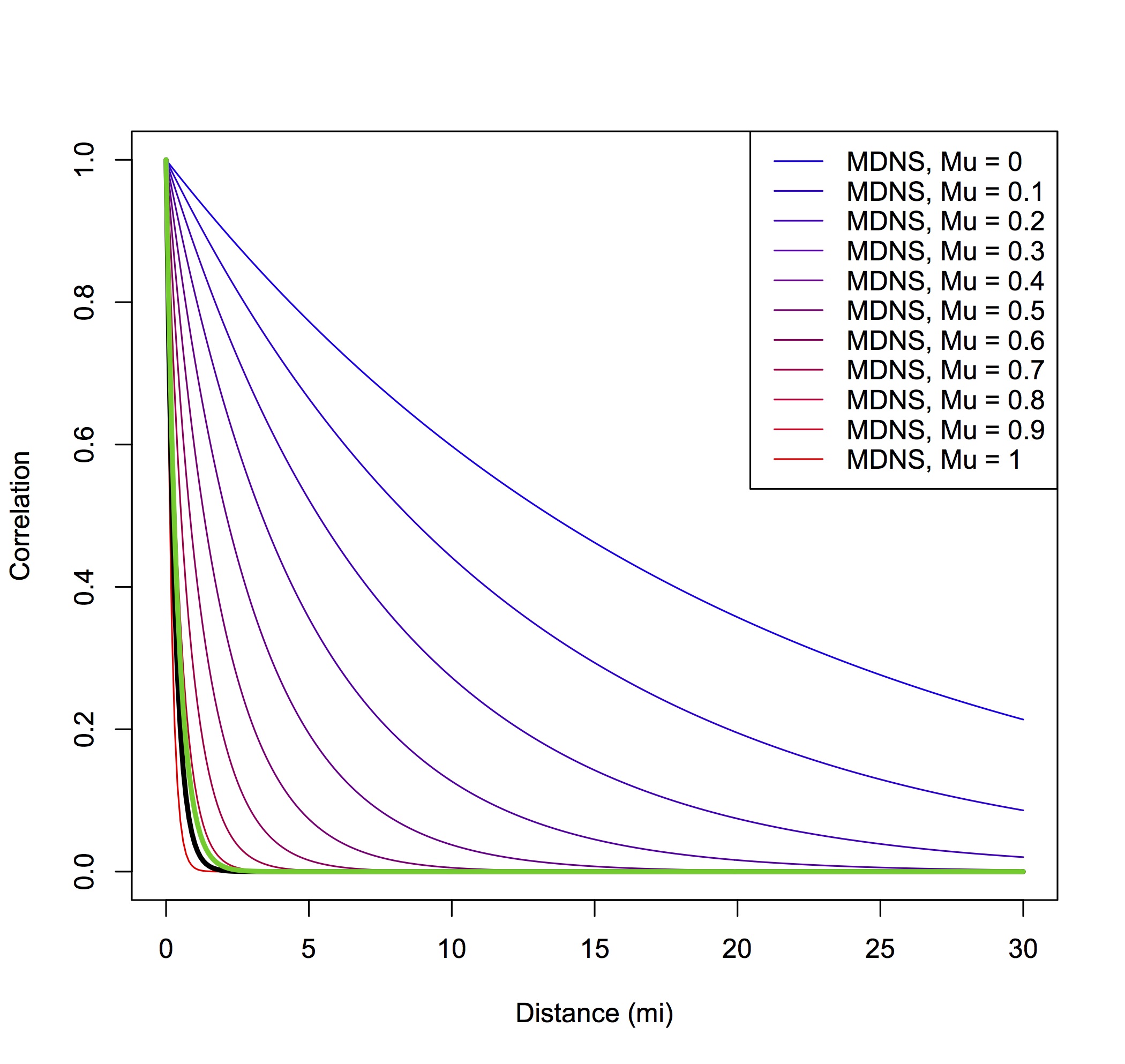}
\includegraphics[height=0.3\textheight]{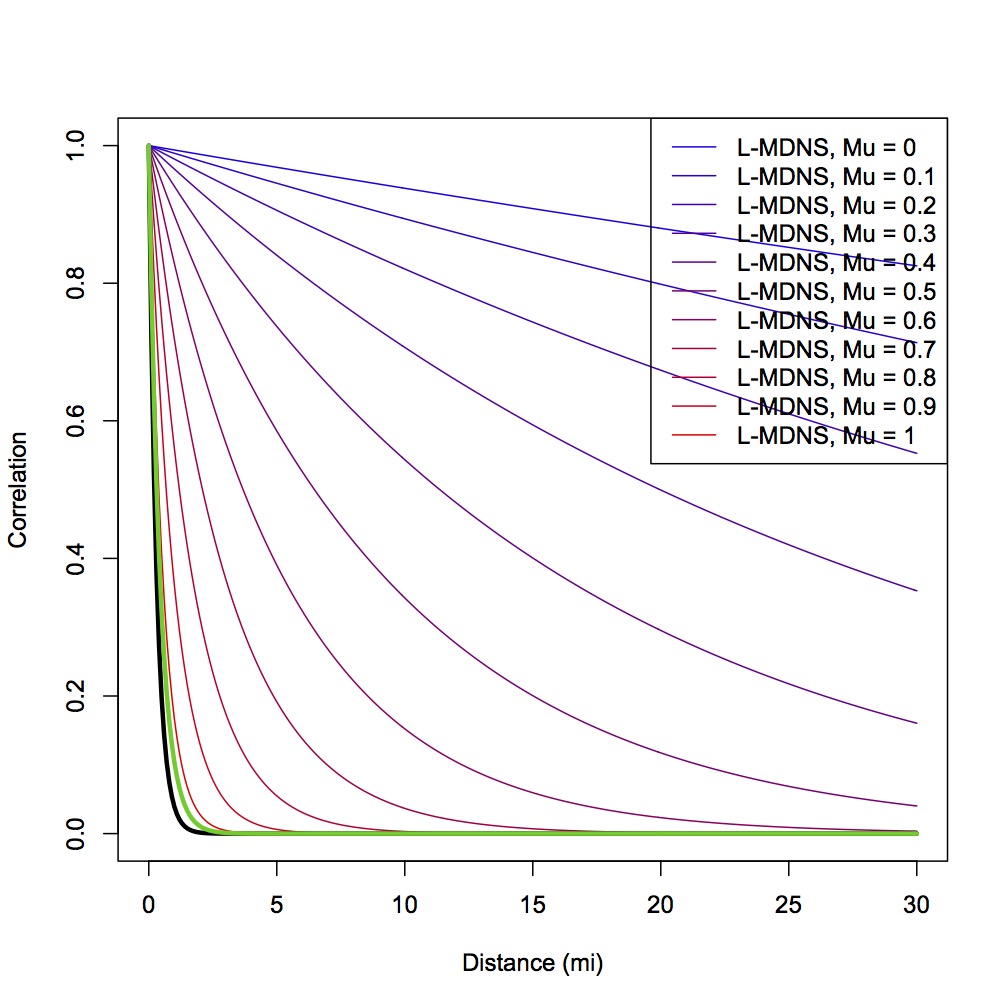}
\caption{Fitted correlation functions for MDNS (left) and L-MDNS (right) for January 2013. Black line is the stationary model fit, green line is the stationary range fits, and the red-to-blue colored lines are the mean-dependent range fits. Mean is fitted with $p=7$ predictors.}
\label{fig:Jan13corrPlots}
\end{center}
\end{figure}

From the parameter estimates in Table \ref{tbl:PR_eta}(a) for the January 2013 data, we observe that the best fitting model is the full MDNS model with $p=7$ predictors. The highly negative values for $a_1$ and $b_1$ indicate that very little nugget effect is present. The nonstationary models are significantly better fits than the stationary model ($\chi^2\geq139.6$ for all $p=4$ fits and $\chi^2\geq87.3$ for all $p=7$ fits). The positive $b_2$ values and negative $b_3$ value indicate that, as the mean increases, the covariability increases and the range of the spatial correlation decreases, as depicted in Figure \ref{fig:Jan13corrPlots}. 

The cross-validation results in Table \ref{tbl:PR_CV}(a) indicate that the predictions from the full MDNS model with $p=7$ predictors are the most accurate in terms of low MSE and high prediction score. Also, the small quantiles for the standard error indicate high precision in the predictions. The coverages indicate that the 95\% prediction intervals of the fully stationary model fail contain all of the removed values. 

\begin{figure}[p]
\centering
\includegraphics[width=0.35\textwidth]{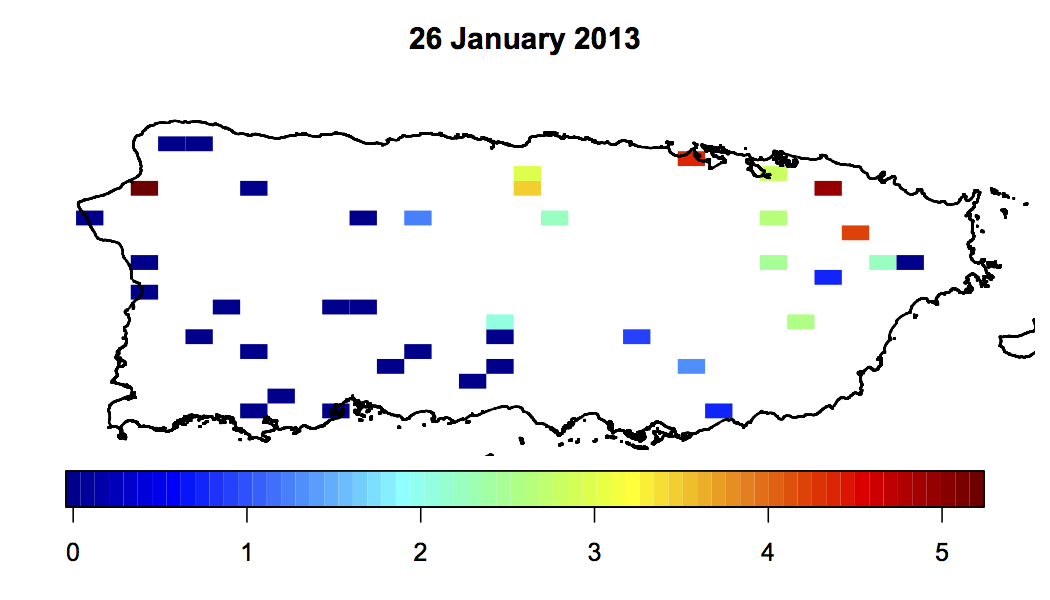}
\includegraphics[width=0.35\textwidth]{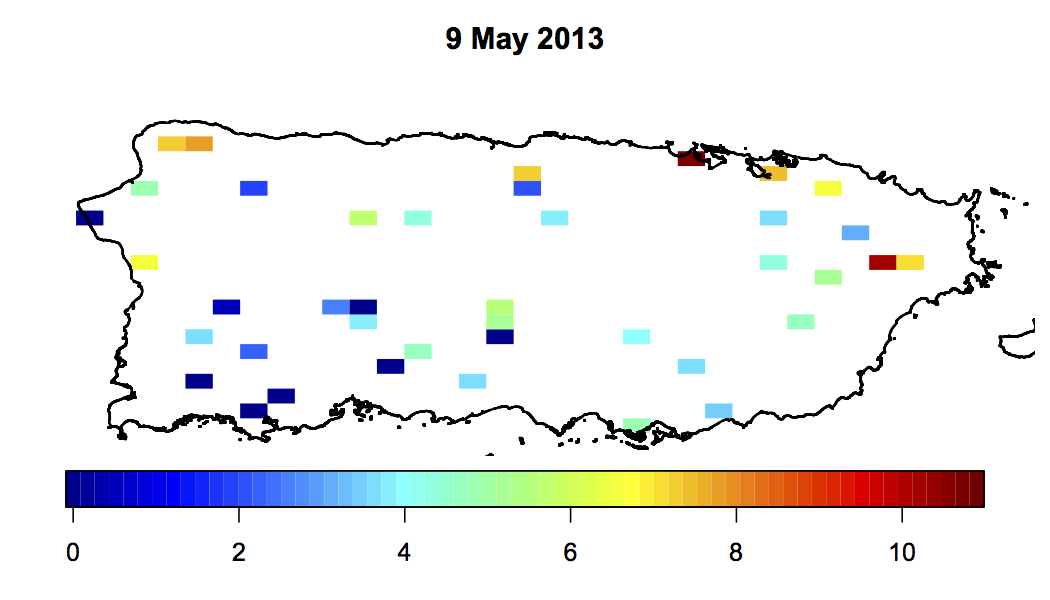}
\caption{Maps of observed values for Puerto Rico data. Rain gauge measurements (in millimeters) obtained from weather stations in Puerto Rico on 26 January 2013 (left) and on 9 May 2013 (right). Data has been square-rooted.}
\label{fig:dry_rainy_daily}
\end{figure}

\begin{figure}[p]
\begin{center}
\includegraphics[height=0.25\textheight]{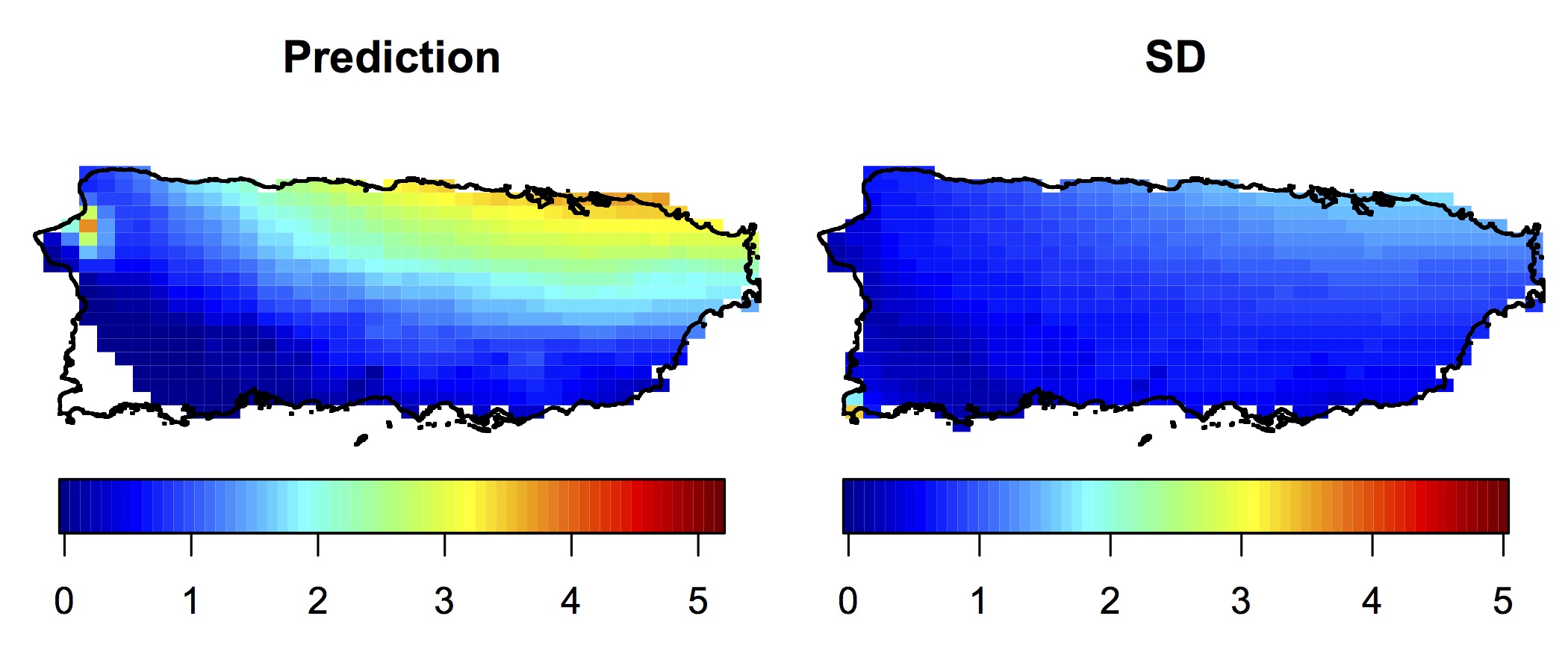}\\
\caption{Predictions and standard deviations for 26 Jan. 2013 using full MDNS and $p=7$ spatial predictors.}
\label{fig:Jan13P7mdns}
\end{center}
\end{figure}

Figure \ref{fig:Jan13P7mdns} depicts the predicted values and standard errors on 26 January 2013, using MDNS and $p=7$ spatial predictors. The raw data appears in Figure \ref{fig:dry_rainy_daily}. The maps predict high precipitation in the northeastern half of the island and practically zero precipitation in the southwestern half, with a possible outlier in the northwest corner. The gaps in the maps indicate regions where the predicted precipitation was below zero, which are treated as zero values for practical purposes.

\subsection{May 2013 data}
\begin{figure}[p]
\begin{center}
\includegraphics[height=0.3\textheight]{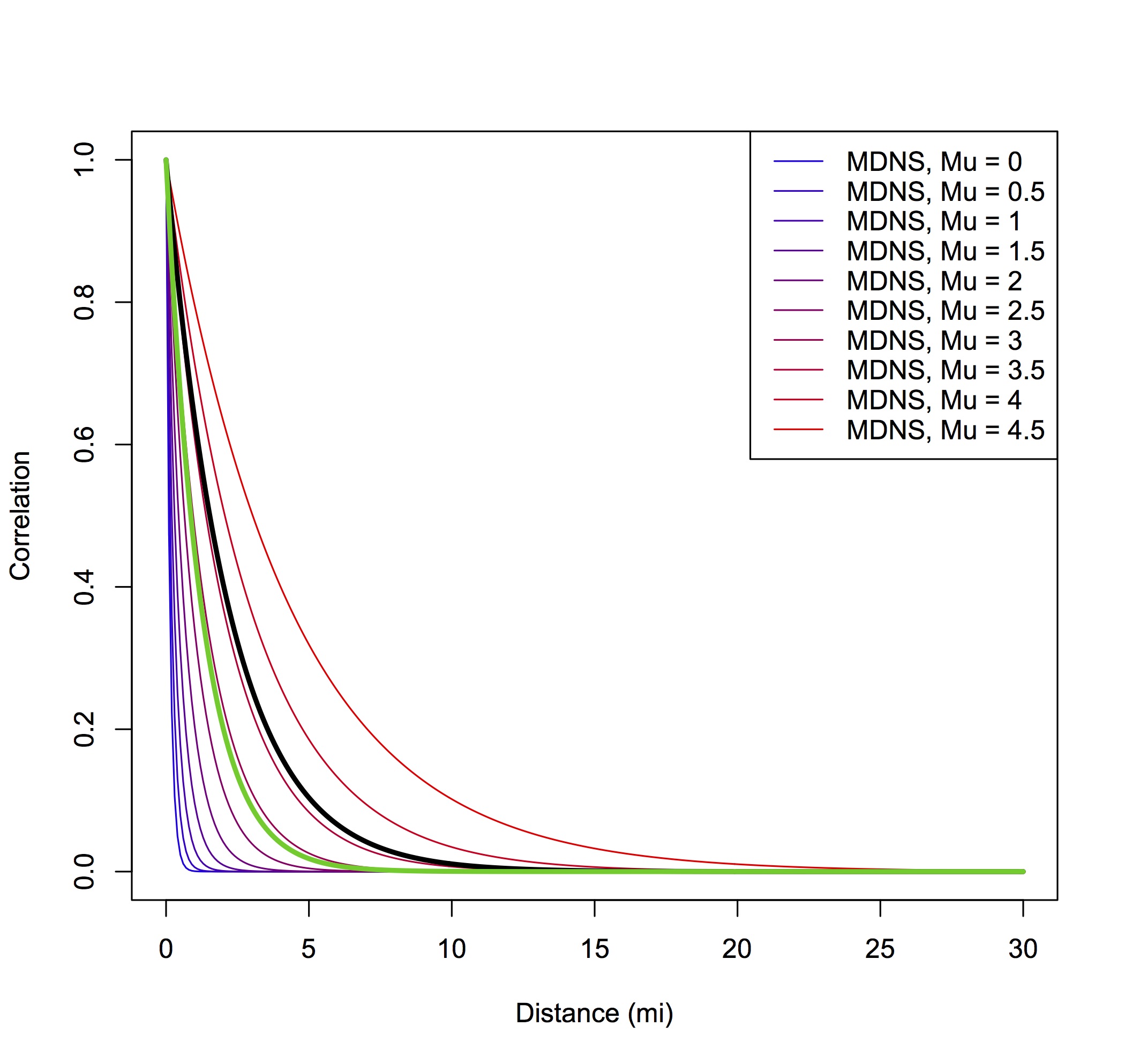}
\includegraphics[height=0.3\textheight]{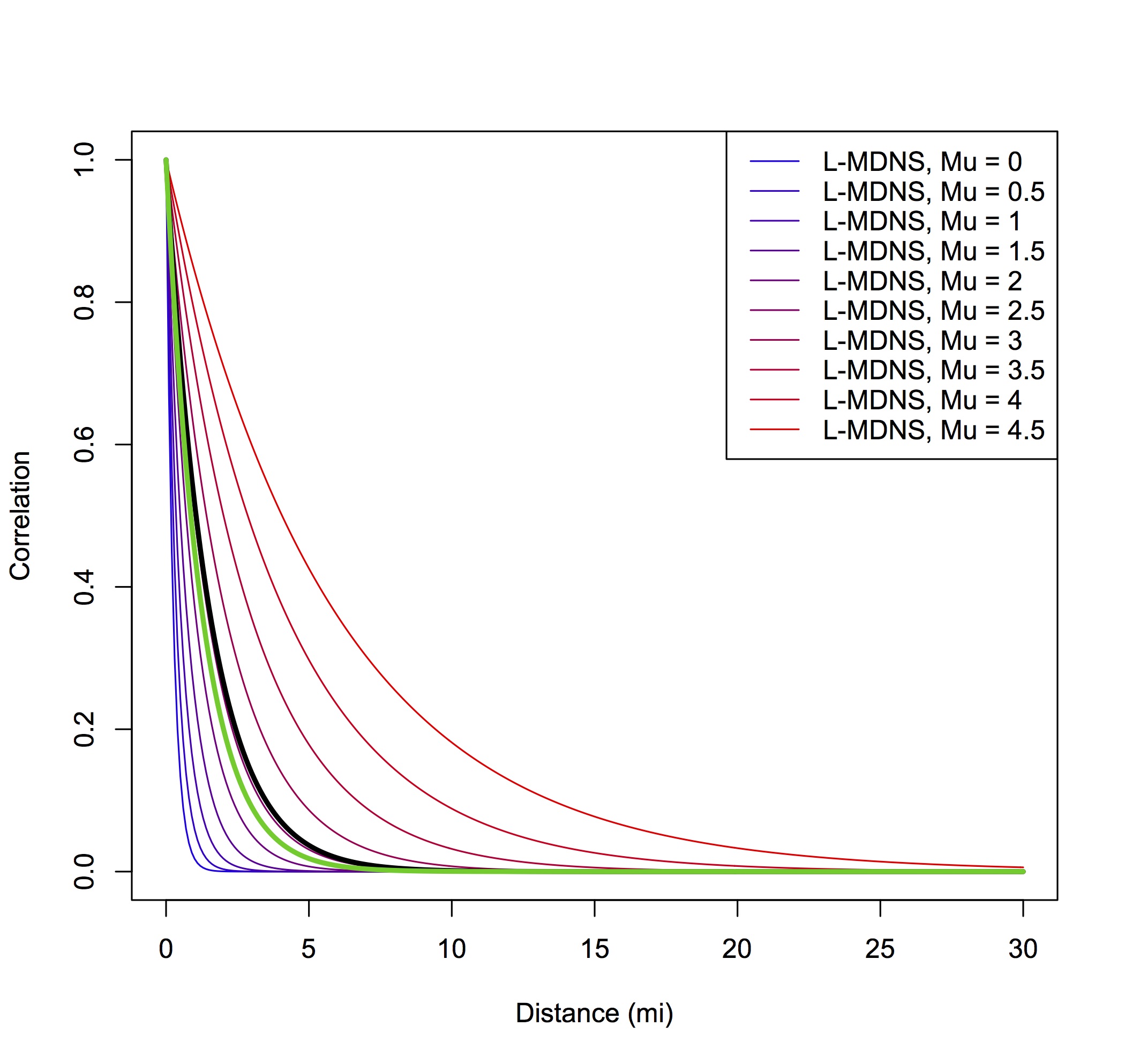}
\caption{Fitted correlation functions for MDNS (left) and L-MDNS (right) for May 2013. Black line is the stationary model fit, green line is the stationary range fits, and the red-to-blue colored lines are the mean-dependent range fits. Mean is fitted with $p=4$ predictors.}
\label{fig:May13corrPlots}
\end{center}
\end{figure}

From the parameter estimates in Table \ref{tbl:PR_eta}(b) for the May 2013 data, we see that the L-MDNS model with $p=4$ predictors is the best fitting model. The nonstationary models are significantly better fits than the stationary model ($\chi^2\geq57.2$ for all $p=4$ fits and $\chi^2\geq76.6$ for all $p=7$ fits). For the range, the difference in log-likelihood between stationary and nonstationary fits are not very significant. The positive values for $b_1$, $b_2$, and $b_3$ indicate that the variability and range increases with the mean, as illustrated in Figure \ref{fig:May13corrPlots}. The highly negative $a_2$ and $b_2$ values for the MDNS model with $p=7$ predictors indicate insignificant spatial covariance in the fitted model, and the highly negative $a_3$ value and positive $b_3$ for the L-MDNS model with $p=7$ predictors indicate that the spatial covariance is significant in areas with high mean precipitation.

From the cross-validation results in Table \ref{tbl:PR_CV}(b), we see that the predictions from full L-MDNS model with $p=7$ predictors are more accurate in terms of low MSE and high prediction score.  However, there are insignificant differences in predictions between the stationary and nonstationary range fits. The coverages indicate that the 95\% prediction intervals of the fully stationary model fail contain all of the removed values.

\begin{figure}[p]
\begin{center}
\includegraphics[height=0.25\textheight]{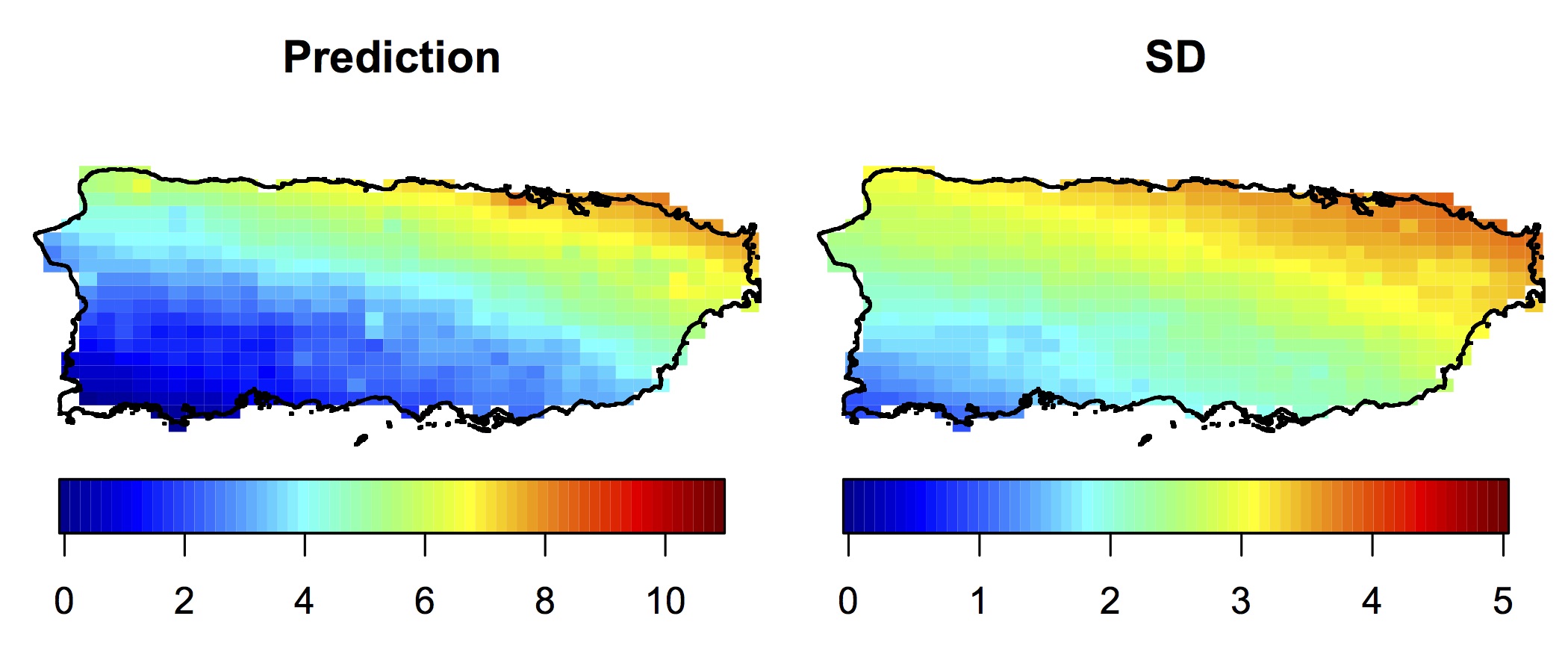}\\
\caption{Predictions and standard deviations for 9 May 2013 using L-MDNS with stationary range and $p=4$ spatial predictors.}
\label{fig:May13P4lmdns}
\end{center}
\end{figure}

Figure \ref{fig:May13P4lmdns} depicts the predicted values and standard errors on 9 May 2013, using L-MDNS with $p=4$ spatial predictors. The raw data appears in Figure \ref{fig:dry_rainy_daily}. From the observed values, there was higher precipitation along the northern coastline and around the eastern Pico El Yunque mountain and lower precipitation in the southeastern coastal plains and central mountains. The predictions depict these features quite well, while the high standard errors indicate that most of the uncertainty lies in the transitory areas between the northern coast and the central mountains. 

\section{Conclusion}
From the simulation study, we see that mean-dependent nonstationary covariance models significantly improve predictions if the data are nonstationary. Testing for nonstationarity appears to be best suited for the single step least-squares method, which is faster to estimate and converges quicker to the asymptotic distribution, and if the test is significant, the full MLE method should be used to compute the most accurate predictions. Using the full MLE for the nonstationarity test is more likely to favor the nonstationary model over the stationary model, but this is not a significant practical problem with the method due to the nonstationary model's overall better predictive fit. Ideally, the full MLE should be used for the test, but the computation cost would have to be drastically reduced through more complex computing methods, such as \cite{stein2004} or \cite{vecchia1988}.

From the Puerto Rico precipitation data analysis, the nonstationary model significantly improves the accuracy and precision of the prediction distribution over the stationary model. For the dry season month of January 2013, the fully mean-dependent model with quadratic spatial predictors created the best prediction distribution. However, assuming local normality while the dry season data has a preponderance of zeroes may result in an ill-behaved likelihood and unrealistic predictions. Future work could include zero-censored normal distributions to address this issue, though this model was designed to be flexible enough to reasonably model both wet and dry seasons without having to use the computation costs of calculating zero-censored multivariate distributions. For the wet season month of May 2013, the fully log-mean-dependent model with quadratic predictors creates the best prediction distributions, but no significant spatial covariance is presence in the fitted model. The log-mean-dependent model with linear predictors has better internal verifiability in terms of log-likelihood, and the spatial covariance is more clearly present in the fitted model.

The results agree with the notion that precipitation measurements are more volatile during heavy rain events. Accounting for this volatility improves prediction and uncertainty quantification, particularly in regards to the possibility of extreme values in areas with high mean values. Improving prediction distributions will be a significant benefit in weather forecasting, reanalysis models, and for developing long-term time series that blend remotely sensed (satellite or radar) and land-based records, particularly for regions with highly varied climates like Puerto Rico. Further research would incorporate temporal or autoregressive effects to predict the mean process for the future days from previous climate data, then utilize the mean- or log-mean-dependent nonstationary spatial model to fully predict the possible range of precipitation values.

\section{Disclaimer}
Use of trade, firm, or product names is for descriptive purposes only and does not imply endorsement by the U.S. government.
\bibliography{pr_arxiv}
\clearpage
\clearpage
\clearpage
\clearpage
\begin{table}[p]
\caption{Estimation Accuracy for Simulated Data: Mean squared error (with standard errors) for covariance parameter estimates, averaged over all simulations. Values are totaled over the 6 elements in $\mbox{\boldmath $\eta$}$ and multiplied by 1000 for clarity. See Appendix \ref{app:etaMSEs} for MSEs for individual elements.}
\begin{center}
\begin{tabular}{|c|c|c||r|r|r|}
  \hline
\multicolumn{3}{|c||}{Training Data} & \multicolumn{3}{|c|}{Models} \\ \hline
 True Covariance & Sites & Days & Stationary & One-step & Full MLE \\ \hline
 \multirow{6}{*}{Stationary} & \multirow{2}{*}{50} & 5 & 218.3 (1.0) & 1241.6 (36.1) & 1330.2 (34.5) \\ 
  & & 10 & 198.1 (0.6) & 311.9 (4.6) & 312.4 (4.5) \\  
  & \multirow{2}{*}{100} & 5 & 96.5 (0.4) & 113.2 (0.5) & 114.0 (0.5) \\ 
  & & 10 & 87.0 (0.3) & 95.6 (0.3) & 95.2 (0.3) \\ 
  & \multirow{2}{*}{200} & 5 & 56.0 (0.3) & 65.6 (0.3) & 65.8 (0.3) \\ 
  & & 10 & 52.5 (0.2) & 56.3 (0.2) & 56.2 (0.2) \\ \hline
  \multirow{6}{*}{Nonstationary} & \multirow{2}{*}{50} & 5 & 1628.9 (4.6) & 683.7 (8.8) & 1767.6 (14.6) \\ 
  & & 10 & 1304.2 (2.5) & 502.6 (1.2) & 1735.7 (10.9) \\ 
  & \multirow{2}{*}{100} & 5 & 746.2 (2.1) & 147.4 (0.7) & 182.8 (0.9) \\ 
  & & 10 & 541.7 (1.0) & 99.0 (0.4) & 108.9 (0.5) \\ 
  & \multirow{2}{*}{200} & 5 & 1133.2 (2.0) & 78.4 (0.4) & 69.6 (0.5) \\ 
  & & 10 & 868.2 (0.9) & 56.1 (0.2) & 34.7 (0.2) \\ \hline
\end{tabular}
\end{center}
\label{tbl:EtaEst}
\end{table}
\begin{table}[p]
\caption{Mean computation time (in seconds) for simulated data.}
\begin{center}
\begin{tabular}{|c|c|c||r|r|r|}
  \hline
\multicolumn{3}{|c||}{Training Data} & \multicolumn{3}{|c|}{Models} \\ \hline
 True Covariance & Sites & Days & Stationary & One-step & Full MLE \\ \hline
\multirow{6}{*}{Stationary} & \multirow{2}{*}{50} & 5 & 8.9 & 18.7 & 81.3 \\ 
& & 10 & 18.3 & 38.8 & 231.8 \\  
& \multirow{2}{*}{100} & 5 & 27.8 & 61.2 & 172.4 \\ 
& & 10 & 54.3 & 112.5 & 443.4 \\  
& \multirow{2}{*}{200} & 5 &132.4 & 264.3 & 686.3 \\ 
& & 10 & 241.6 & 444.9 & 1442.0 \\ \hline
\multirow{6}{*}{Nonstationary} & \multirow{2}{*}{50} & 5 & 8.0 & 15.6 & 219.0 \\ 
& & 10 & 15.7 & 27.5 & 1186.7 \\  
& \multirow{2}{*}{100} & 5 & 25.3 & 71.2 & 531.6 \\ 
& & 10 & 53.0 & 146.9 & 2919.3 \\  
& \multirow{2}{*}{200} & 5 & 109.6 & 328.3 & 2128.6 \\ 
& & 10 & 242.0 & 665.0 & 11589.0 \\ \hline
\end{tabular}
\end{center}
\label{tbl:Times}
\end{table}
\clearpage
\begin{table}[p]
\caption{Evaluation of Prediction Distributions for Simulated Data: Percent improvement in mean squared errors (top) and differences in deviance (bottom) between the nonstationary and stationary models. Positive percentages indicate that the nonstationary model was better than the stationary model. Values in parentheses are standard errors.}
(a) Percent Difference in Mean Squared Errors (MSE)\\
\begin{center}
\begin{tabular}{|c|c|c||r|r|}
  \hline
\multicolumn{3}{|c||}{Training Data} & \multicolumn{2}{|c|}{Models} \\ \hline
True Covariance & Sites & Days & One-step & Full MLE \\ \hline
\multirow{6}{*}{Stationary} & \multirow{2}{*}{50} & 5 & -0.55 (0.12) & -0.82 (0.15) \\ 
& & 10 & -0.22 (0.04) & -0.31 (0.07) \\ 
& \multirow{2}{*}{100} & 5 & -0.26 (0.05) & -0.27 (0.06) \\ 
& & 10 & -0.14 (0.02) & -0.16 (0.03) \\ 
& \multirow{2}{*}{200} & 5 & -0.16 (0.04) & -0.17 (0.04) \\ 
& & 10 & -0.07 (0.02) & -0.07 (0.02) \\ \hline
\multirow{6}{*}{Nonstationary} & \multirow{2}{*}{50} & 5 & 1.04 (0.20) & 1.24 (0.37) \\ 
& & 10 & 0.98 (0.13) & 1.74 (0.24) \\ 
& \multirow{2}{*}{100} & 5 & 1.12 (0.26) & 2.44 (0.02) \\ 
& & 10 & 1.32 (0.17) & 2.89 (0.23) \\ 
& \multirow{2}{*}{200} & 5 & 1.54 (0.42) & 2.28 (0.44) \\ 
& & 10 & 2.10 (0.24) & 3.05 (0.27) \\ \hline
\end{tabular}\\
\end{center}
(b) Difference in Deviance\\
\begin{center}
\begin{tabular}{|c|c|c||r|r|}
  \hline
\multicolumn{3}{|c||}{Training Data} & \multicolumn{2}{|c|}{Models} \\ \hline
True Covariance & Sites & Days & \multicolumn{1}{|c|}{One-step LS} & Full MLE \\ \hline
\multirow{6}{*}{Stationary} & \multirow{2}{*}{50} & 5 & -10.31 (1.40) & -13.05 (1.74) \\ 
& & 10 & -11.83 (1.69) & -13.15 (1.83) \\ 
&  \multirow{2}{*}{100} & 5 & -4.05 (0.43) & -4.29 (0.46) \\ 
& & 10 & -4.51 (0.40) & -4.90 (0.44) \\ 
&  \multirow{2}{*}{200} & 5 & -1.80 (0.24) & -1.97 (0.25) \\ 
& & 10 & -1.58 (0.22) & -1.68 (0.23) \\ \hline
\multirow{6}{*}{Nonstationary} & \multirow{2}{*}{50} & 5 & 250.53 (7.13) & 280.86 (7.10) \\ 
& & 10 & 420.61 (9.14) & 483.65 (9.03) \\ 
&  \multirow{2}{*}{100} & 5 & 249.03 (4.59) & 265.79 (4.51) \\ 
& & 10 & 450.30 (6.00) & 484.79 (6.02) \\ 
&  \multirow{2}{*}{200} & 5 & 294.04 (5.08) & 305.37 (5.19) \\ 
& & 10 & 519.41 (6.36) & 543.94 (6.50) \\ \hline
\end{tabular}
\end{center}
\label{tbl:PredMSEScore}
\end{table}
\clearpage
\begin{table}[p]
\caption{Covariance parameter estimates and log-likelihoods for Puerto Rico precipitation data. The nugget and sill are fitted under stationary, mean-dependent nonstationarity (MDNS), or log-mean-dependent nonstationarity (L-MDNS) conditions. Range is fitted under stationary or mean-dependent nonstationarity (MDNS) conditions.}
(a) January 2013\\
\begin{center}
\begin{tabular}{|c|c|c||c||r|r|r|r|r|r|}\hline
\multicolumn{3}{|c||}{Model} & & \multicolumn{2}{|c|}{Nugget} & \multicolumn{2}{|c|}{Sill} & \multicolumn{2}{|c|}{Range} \\ \hline
$p$ & Nugget/Sill & Range & Log-Likelihood & $a_1$ & $b_1$ & $a_2$ & $b_2$ & $a_3$ & $b_3$ \\ \hline
\multirow{5}{*}{4} & Stationary & Stationary & -1575.7 & -0.91 & 0 & -2.22 & 0 & 2.70 & 0 \\
 & MDNS & Stationary & -1436.1 & -1.72 & 1.08 & -5.24 & -9.27 & 54.78 & 0 \\ 
 & MDNS & MDNS & -1431.9 & -1.80 & 1.15 & -4.26 & -7.34 & 48.20 & 11.18 \\
 & LMDNS & Stationary & -1435.2 & -2.56 & 1.95 & -1.80 & 0.72 & 1.88 & 0 \\ 
 & LMDNS & MDNS & -1421.2 & -2.60 & 1.99 & -4.62 & 8.94 & 2.62 & -12.15 \\ \hline
\multirow{5}{*}{7} & Stationary & Stationary & -1580.1 & -1.48 & 0 & -1.52 & 0 & -2.35 & 0 \\
 & MDNS & Stationary & -1464.1 & -7.55 & -11.09 & -1.70 & 0.92 & -1.80 & 0 \\ 
 & MDNS & MDNS & -1405.9 & -17.79 & -24.30 & -1.36 & 0.65 & 5.93 & -9.27 \\
 & LMDNS & Stationary & -1492.8 & -2.36 & 1.45 & -0.93 & 1.11 & -1.64 & 0 \\ 
 & LMDNS & MDNS & -1452.1 & -2.27 & 1.37 & 3.03 & 4.29 & 10.10 & -11.27 \\ 
\hline
\end{tabular}\\
\end{center}
(b) May 2013\\
\begin{center}
\begin{tabular}{|c|c|c||c||r|r|r|r|r|r|}\hline
\multicolumn{3}{|c||}{Model} & & \multicolumn{2}{|c|}{Nugget} & \multicolumn{2}{|c|}{Sill} & \multicolumn{2}{|c|}{Range} \\ \hline
$p$ & Nugget/Sill & Range & Log-Likelihood & $a_1$ & $b_1$ & $a_2$ & $b_2$ & $a_3$ & $b_3$ \\ \hline
\multirow{5}{*}{4} & Stationary & Stationary & -2774.5 & 0.94 & 0 & 0.08 & 0 & 1.58 & 0 \\
 & MDNS & Stationary & -2717.3 & -0.09 & 0.34 & -0.26 & 0.35 & 0.44 & 0 \\ 
 & MDNS & MDNS & -2713.7 & -3.14 & 0.34 & 0.42 & 0.39 & -4.03 & 1.55 \\
 & LMDNS & Stationary & -2693.9 & -0.78 & 1.27 & 0.38 & 3.67 & 0.45 & 0 \\ 
 & LMDNS & MDNS & -2692.4 & -0.81 & 1.30 & -0.49 & 4.31 & -2.79 & 1.41 \\ \hline
\multirow{5}{*}{7} & Stationary & Stationary & -2830.9 & 0.22 & 0 & 0.75 & 0 & -3.77 & 0 \\
 & MDNS & Stationary & -2754.3 & 0.32 & 0.36 & -2.08 & -2.76 & 0.83 & 0 \\ 
 & MDNS & MDNS & -2747.5 & 0.35 & 0.35 & -2.19 & -2.37 & 23.70 & -4.02 \\
 & LMDNS & Stationary & -2753.7 & -0.52 & 0.94 & -0.43 & 4.95 & -3.73 & 0 \\ 
 & LMDNS & MDNS & -2753.2 & -0.52 & 0.94 & 7.21 & 2.29 & -15.82 & 3.20 \\ 
\hline
\end{tabular}
\end{center}
\label{tbl:PR_eta}
\end{table}
\begin{table}[p]
\caption{Five-fold cross-validation results for Puerto Rico precipitation data. For predictive accuracy, we calculate the score, mean-squared error, and coverage of the 95\% prediction interval using the prediction distribution. The quantiles of the standard errors for the test sites illustrate how the nonstationary models compare to the stationary models in terms of precision.}
(a) January 2013\\
\begin{center}
\begin{tabular}{|c|c|c||c|c|c|c|c|c|}\hline
\multicolumn{3}{|c||}{Model} & \multicolumn{3}{|c|}{Prediction Accuracy} & \multicolumn{3}{|c|}{Standard Errors} \\\hline $p$ & Nugget/Sill & Range & Score & MSE & Coverage & $Q_{.05}$ & $Q_{.50}$ & $Q_{.95}$ \\ \hline
\multirow{5}{*}{4} & Stationary & Stationary & -2.555 & 0.518 & 93.5\% & 0.69 & 0.72 & 0.72 \\ 
 & MDNS & Stationary & -0.335 & 0.505 & 96.0\% & 0.38 & 0.59 & 1.23 \\ 
 & MDNS & MDNS & -0.431 & 0.508 & 95.7\% & 0.39 & 0.58 & 1.27 \\
 & LMDNS & Stationary & -0.424 & 0.513 & 95.8\% & 0.35 & 0.62 & 1.12 \\ 
 & LMDNS & MDNS & -0.060 & 0.505 & 96.3\% & 0.33 & 0.62 & 1.13 \\ \hline
\multirow{5}{*}{7} & Stationary & Stationary & -1.305 & 0.451 & 94.8\% & 0.64 & 0.67 & 0.69 \\ 
 & MDNS & Stationary & +0.388 & 0.450 & 96.2\% & 0.39 & 0.57 & 1.14 \\ 
 & MDNS & MDNS & +1.715 & 0.441 & 95.3\% & 0.22 & 0.61 & 1.02 \\
 & LMDNS & Stationary & -0.264 & 0.453 & 96.0\% & 0.37 & 0.62 & 1.00 \\ 
 & LMDNS & MDNS & +0.546 & 0.444 & 95.9\% & 0.34 & 0.62 & 1.00 \\ \hline
\end{tabular}\\
\end{center}
(b) May 2013\\
\begin{center}
\begin{tabular}{|c|c|c||c|c|c|c|c|c|}\hline
\multicolumn{3}{|c||}{Model} & \multicolumn{3}{|c|}{Prediction Accuracy} & \multicolumn{3}{|c|}{Standard Errors} \\\hline $p$ & Nugget/Sill & Range & Score & MSE & Coverage & $Q_{.05}$ & $Q_{.50}$ & $Q_{.95}$ \\ \hline
\multirow{5}{*}{4} & Stationary & Stationary & -20.21 & 3.655 & 94.8\% & 1.86 & 1.91 & 1.93 \\ 
 & MDNS & Stationary & -19.33 & 3.645 & 95.1\% & 1.30 & 1.71 & 2.90 \\ 
 & MDNS & MDNS & -19.20 & 3.676 & 94.9\% & 1.27 & 1.71 & 2.75 \\
 & LMDNS & Stationary & -19.04 & 3.622 & 95.4\% & 1.04 & 1.76 & 2.87 \\ 
 & LMDNS & MDNS & -19.07 & 3.676 & 95.4\% & 1.02 & 1.77 & 2.80 \\ \hline
\multirow{5}{*}{7} & Stationary & Stationary & -19.39 & 3.335 & 94.5\% & 1.81 & 1.82 & 1.86 \\ 
 & MDNS & Stationary & -18.68 & 3.315 & 95.2\% & 1.19 & 1.61 & 3.07 \\ 
 & MDNS & MDNS & -18.53 & 3.296 & 95.2\% & 1.19 & 1.61 & 3.07 \\
 & LMDNS & Stationary & -18.19 & 3.335 & 95.2\% & 1.11 & 1.74 & 2.54 \\ 
 & LMDNS & MDNS & -18.18 & 3.342 & 95.2\% & 1.10 & 1.74 & 2.52 \\ \hline
\end{tabular}
\end{center}
\label{tbl:PR_CV}
\end{table}
\clearpage
\clearpage

\clearpage

\newpage
\appendix
\section{Covariance parameter MSEs from Simulation Study}\label{app:etaMSEs}
\begin{table}[ht]
Averaged squared error (and standard errors) for $\eta_1=\exp(a_1)$ estimates. (Multiplied by 10000 for clarity.)\\
\begin{center}
\begin{tabular}{|c|c|c||r|r|r|}
  \hline
\multicolumn{3}{|c||}{Training Data} & \multicolumn{3}{|c|}{Models} \\ \hline
True Covariance & Sites & Days & Stationary& One-step & Full MLE \\ \hline
\multirow{6}{*}{Stationary} & \multirow{2}{*}{50} & 5 & 654.2 (2.8) & 644.6 (2.8) & 645.2 (2.9) \\ 
& & 10 & 496.1 (2.3) & 516.3 (2.4) & 499.2 (2.4) \\  
& \multirow{2}{*}{100} & 5 & 111.0 (0.7) & 155.5 (1.0) & 154.3 (1.0) \\ 
& & 10 & 67.9 (0.5) & 99.1 (0.6) & 98.9 (0.6) \\  
& \multirow{2}{*}{200} & 5 & 57.9 (0.4) & 72.2 (0.5) & 72.2 (0.5) \\ 
& & 10 & 33.6 (0.2) & 37.7 (0.3) & 38.1 (0.3) \\ \hline
\multirow{6}{*}{Nonstationary} & \multirow{2}{*}{50} & 5 & 1412.4 (8.3) & 1278.2 (1.6) & 1198.3 (1.9) \\ 
& & 10 & 1313.1 (1.0) & 1294.8 (1.3) & 1189.6 (1.7) \\  
& \multirow{2}{*}{100} & 5 & 1454.8 (9.6) & 210.0 (1.5) & 199.3 (1.2) \\ 
& & 10 & 916.7 (5.5) & 111.3 (0.7) & 103.3 (0.7) \\  
& \multirow{2}{*}{200} & 5 & 313.3 (2.2) & 92.1 (0.7) & 70.2 (0.6) \\ 
& & 10 & 167.9 (1.5) & 47.1 (0.3) & 29.1 (0.2) \\ \hline
\end{tabular}\\
\end{center}
Averaged squared error (and standard errors) for $\eta_2=b_1$ estimate. (Multiplied by 10000 for clarity.)\\
\begin{center}
\begin{tabular}{|c|c|c||r|r|r|}
  \hline
\multicolumn{3}{|c||}{Training Data} & \multicolumn{3}{|c|}{Models} \\ \hline
True Covariance & Sites & Days & Stationary & One-step & Full MLE \\ \hline
\multirow{6}{*}{Stationary} & \multirow{2}{*}{50} & 5 & 0.0 (0.0) & 59826.4 (2152.2) & 64434.2 (2050.3) \\ 
& & 10 & 0.0 (0.0) & 6319.1 (269.7) & 6362.8 (266.0) \\  
& \multirow{2}{*}{100} & 5 & 0.0 (0.0) & 272.6 (2.2) & 283.4 (2.2) \\ 
& & 10 & 0.0 (0.0) & 156.2 (1.1) & 161.1 (1.1) \\  
& \multirow{2}{*}{200} & 5 & 0.0 (0.0) & 111.6 (0.9) & 117.6 (0.9) \\ 
& & 10 & 0.0 (0.0) & 50.9 (0.3) & 53.3 (0.4) \\ \hline
\multirow{6}{*}{Nonstationary} & \multirow{2}{*}{50} & 5 & 100.0 (0.0) & 14001.3 (517.1) & 64865.8 (667.9) \\ 
& & 10 & 100.0 (0.0) & 5481.6 (38.7) & 79314.5 (643.8) \\  
& \multirow{2}{*}{100} & 5 & 100.0 (0.0) & 1366.0 (11.1) & 1496.6 (28.0) \\ 
& & 10 & 100.0 (0.0) & 776.1 (6.3) & 602.3 (5.9) \\  
& \multirow{2}{*}{50} & 5 & 100.0 (0.0) & 178.5 (2.4) & 183.7 (6.0) \\ 
& & 10 & 100.0 (0.0) & 136.0 (3.0) & 49.2 (1.1) \\ \hline
\end{tabular}
\end{center}
\end{table}

\begin{table}[ht]
Averaged squared error (and standard errors) for $\eta_3=\exp(a_2)$ estimate. (Multiplied by 10000 for clarity.)\\
\begin{center}
\begin{tabular}{|c|c|c||r|r|r|}
  \hline
\multicolumn{3}{|c||}{Training Data} & \multicolumn{3}{|c|}{Models} \\ \hline
True Covariance & Sites & Days & Stationary & One-step & Full MLE \\ \hline
\multirow{6}{*}{Stationary} & \multirow{2}{*}{50} & 5 & 1024.5 (7.9) & 1040.9 (8.0) & 1077.9 (8.3) \\ 
& & 10 & 702.1 (4.9) & 734.7 (5.5) & 780.5 (5.9) \\  
& \multirow{2}{*}{100} & 5 & 699.5 (3.8) & 705.7 (3.8) & 701.9 (3.8) \\ 
& & 10 & 590.7 (3.0) & 568.0 (3.2) & 560.1 (3.1) \\  
& \multirow{2}{*}{200} & 5 & 416.5 (2.5) & 453.6 (2.8) & 450.9 (2.8) \\ 
& & 10 & 421.9 (2.0) & 441.5 (2.1) & 434.4 (2.1) \\ \hline
\multirow{6}{*}{Nonstationary} & \multirow{2}{*}{50} & 5 & 56032.7 (226.5) & 2503.7 (15.0) & 1003.8 (8.3) \\ 
& & 10 & 39556.9 (96.0) & 1841.9 (10.2) & 643.5 (4.4) \\  
& \multirow{2}{*}{100} & 5 & 26599.4 (116.9) & 1156.4 (14.1) & 829.4 (4.8) \\ 
& & 10 & 16148.4 (54.2) & 526.3 (4.9) & 754.9 (4.4) \\  
& \multirow{2}{*}{200} & 5 & 44648.2 (122.4) & 746.8 (7.8) & 772.3 (4.4) \\ 
& & 10 & 30649.4 (54.4) & 319.9 (4.0) & 493.0 (3.0) \\ \hline
\end{tabular}\\
\end{center}
Averaged squared error (and standard errors) for $\eta_4=b_2$ estimate. (Multiplied by 10000 for clarity.)\\
\begin{center}
\begin{tabular}{|c|c|c||r|r|r|}
  \hline
\multicolumn{3}{|c||}{Training Data} & \multicolumn{3}{|c|}{Models} \\ \hline
True Covariance & Sites & Days & Stationary & One-step & Full MLE \\ \hline
\multirow{6}{*}{Stationary} & \multirow{2}{*}{50} & 5 & 0.0 (0.0) & 57.5 (0.5) & 75.9 (0.6) \\ 
& & 10 & 0.0 (0.0) & 34.6 (0.3) & 51.0 (0.5) \\  
& \multirow{2}{*}{100} & 5 & 0.0 (0.0) & 32.3 (0.2) & 38.0 (0.3) \\ 
& & 10 & 0.0 (0.0) & 22.5 (0.1) & 26.3 (0.2) \\  
& \multirow{2}{*}{200} & 5 & 0.0 (0.0) & 22.9 (0.1) & 28.4 (0.2) \\ 
& & 10 & 0.0 (0.0) & 12.7 (0.1) & 14.6 (0.1) \\ \hline
\multirow{6}{*}{Nontationary} & \multirow{2}{*}{50} & 5 & 2500.0 (0.0) & 201.4 (1.2) & 90.1 (0.6) \\ 
& & 10 & 2500.0 (0.0) & 178.5 (0.8) & 93.1 (0.6) \\  
& \multirow{2}{*}{100} & 5 & 2500.0 (0.0) & 239.0 (1.6) & 106.9 (1.2) \\ 
& & 10 & 2500.0 (0.0) & 162.0 (1.2) & 54.1 (0.4) \\  
& \multirow{2}{*}{200} & 5 & 2500.0 (0.0) & 80.9 (0.9) & 40.2 (0.3) \\ 
& & 10 & 2500.0 (0.0) & 49.3 (0.4) & 22.3 (0.2) \\ \hline
\end{tabular}
\end{center}
\end{table}

\begin{table}[ht]
Averaged squared error (and standard errors) for $\eta_5=a_3$ estimate. (Multiplied by 10000 for clarity.)\\
\begin{center}
\begin{tabular}{|c|c|c||r|r|r|}
  \hline
\multicolumn{3}{|c||}{Training Data} & \multicolumn{3}{|c|}{Models} \\ \hline
True Covariance & Sites & Days & Stationary & One-step & Full MLE \\ \hline
\multirow{6}{*}{Stationary} & \multirow{2}{*}{50} & 5 & 11421.5 (57.5) & 12167.9 (65.3) & 12099.2 (67.2) \\ 
& & 10 & 10688.2 (35.1) & 10788.3 (35.3) & 10582.3 (35.5) \\  
& \multirow{2}{*}{100} & 5 & 4981.9 (24.7) & 5290.1 (271) & 5182.4 (27.1) \\ 
& & 10 & 4561.4 (18.8) & 4702.6 (19.3) & 4627.8 (19.4) \\  
& \multirow{2}{*}{200} & 5 & 2887.9 (14.2) & 3108.9 (151) & 3055.3 (15.0) \\ 
& & 10 & 2694.4 (9.3) & 2756.5 (10.2) & 2724.6 (10.3) \\ \hline
\multirow{6}{*}{Nonstationary} & \multirow{2}{*}{50} & 5 & 35189.6 (165.6) & 20653.4 (92.7) & 27312.4 (138.8) \\ 
& & 10 & 32284.0 (116.1) & 20253.4 (63.2) & 20338.2 (69.2) \\  
& \multirow{2}{*}{100} & 5 & 11614.9 (50.7) & 5166.4 (32.1) & 6438.8 (37.8) \\ 
& & 10 & 10337.6 (36.4) & 3954.5 (20.0) & 3939.0 (22.9) \\  
& \multirow{2}{*}{200} & 5 & 17930.9 (39.4) & 3291.7 (18.8) & 2589.1 (20.3) \\ 
& & 10 & 16175.1 (29.4) & 2617.6 (11.4) & 1261.9 (8.9) \\ \hline
\end{tabular}\\
\end{center}
Averaged squared error (and standard errors) for $\eta_6=b_3$ estimate. (Multiplied by 10000 for clarity.)\\
\begin{center}
\begin{tabular}{|c|c|c||r|r|r|}
  \hline
\multicolumn{3}{|c||}{Training Data} & \multicolumn{3}{|c|}{Models} \\ \hline
True Covariance & Sites & Days & Stationary & One-step & Full MLE \\ \hline
\multirow{6}{*}{Stationary} & \multirow{2}{*}{50} & 5 & 0.0 (0.0) & 759.0 (11.3) & 1481.9 (30.1) \\ 
& & 10 & 0.0 (0.0) & 318.6 (3.2) & 467.9 (4.6) \\  
& \multirow{2}{*}{100} & 5 & 0.0 (0.0) & 336.0 (3.7) & 478.6 (6.1) \\ 
& & 10 & 0.0 (0.0) & 190.4 (1.6) & 238.4 (2.0) \\  
& \multirow{2}{*}{200} & 5 & 0.0 (0.0) & 164.2 (1.3) & 226.0 (2.0) \\ 
& & 10 & 0.0 (0.0) & 81.3 (0.6) & 104.7 (0.9) \\ \hline
\multirow{6}{*}{Nonstationary} & \multirow{2}{*}{50} & 5 & 2500.0 (0.0) & 2381.4 (72.2) & 11587.5 (566.3) \\ 
& & 10 & 2500.0 (0.0) & 1108.6 (4.0) & 2565.7 (13.0) \\  
& \multirow{2}{*}{100} & 5 & 2500.0 (0.0) & 707.5 (4.5) & 1898.8 (12.0) \\ 
& & 10 & 2500.0 (0.0) & 412.1 (2.7) & 1080.5 (6.2) \\  
& \multirow{2}{*}{200} & 5 & 2500.0 (0.0) & 314.3 (3.0) & 522.0 (7.2) \\ 
& & 10 & 2500.0 (0.0) & 195.7 (1.9) & 226.8 (1.3) \\ \hline
\end{tabular}
\end{center}
\end{table}
\end{document}